\documentclass[11pt]{article}
\usepackage{cite}
\usepackage{amsmath,amsfonts,amssymb}
\usepackage[small,bf,hang]{caption}
\usepackage{slashed}

\def\hybrid{
        \topmargin -20pt
        \oddsidemargin 0pt
        \headheight 0pt \headsep 0pt
        \textwidth 6.25in 
        \textheight 9.5in 
        \marginparwidth .875in
        \parskip 5pt plus 1pt \jot = 1.5ex}

\hybrid

 \csname
@addtoreset\endcsname{equation}{section}

\def\moth{\mathsurround=0pt}
\newdimen\zo \zo=0pt

\def\tick{\leaders\hrule height 0.5ex depth 0pt \hskip 0.5pt}
\def\upboxfill{$\moth \setbox\zo\hbox{\tick}%
  \hskip 3pt\hbox to 0pt{$\tick$\hss}\hrulefill \hbox to 7.5pt{$\tick$\hss}$}

\def\dtick{\leaders\hrule height .34pt depth 0.5ex \hskip 0.5pt}
\def\downboxfill{$\moth \setbox\zo\hbox{\dtick}%
  \hskip 2pt\hbox to 0pt{$\dtick$\hss}\hrulefill \hbox to 2pt{$\dtick$\hss}$}

\def\ciftS{\mathbb{S}}
\def\cL{{\cal L}}
\def\cH{{\cal H}}
\def\cK{{\cal K}}
\def\cR{{\cal R}}

\def\hl{\hat\lambda}

\def\hdelta{\hat{\delta}}
\def\hx{\hat{\xi}}

\def\hl{\hat\lambda}

\def\del{\partial}

\def\be{\begin{equation}}
\def\ee{\end{equation}}
\def\bea{\begin{eqnarray}}
\def\eea{\end{eqnarray}}
\def\ba{\begin{array}}
\def\ea{\end{array}}

\begin{document}

\begin{titlepage}
\rightline{} \rightline\today
\begin{center}
\vskip 2.5cm {\Large \bf { Duality Twisted Reductions of Double Field Theory of Type II Strings}}\\
\vskip 2.5cm {\large {Aybike \c{C}atal-\"{O}zer}} \vskip 1cm
{\it {Department of Mathematics,}}\\
{\it {\.{I}stanbul Technical University,}}\\
{\it {Maslak 34469,
\.{I}stanbul, Turkey}}\\
ozerayb@itu.edu.tr \vskip 0.5cm

\vskip 1cm {\bf Abstract}
\end{center}

\vskip 0.5cm

\noindent
\begin{narrower}

\noindent We study  duality twisted reductions of the Double Field
Theory (DFT) of the RR sector of massless Type II theory, with
twists belonging to the duality group $Spin^+(10,10)$. We
determine the action and the gauge algebra of the resulting theory
and determine the conditions for consistency. In doing this, we
work with the DFT action constructed by Hohm, Kwak and Zwiebach,
which we rewrite in terms of the Mukai pairing: a natural bilinear
form on the space of spinors, which is manifestly $Spin(n,n)$
invariant.   If the duality twist is introduced via the
$Spin^+(10,10)$ element $S$ in the RR sector, then the NS-NS
sector should also  be deformed via the duality twist
 $U = \rho(S)$, where $\rho$ is the double
covering homomorphism between $Pin(n,n)$ and $O(n,n)$.  We show
that the set of conditions required for the consistency of the
reduction of the NS-NS sector are also crucial for the consistency
of the reduction of the RR sector, owing to the fact that  the Lie
algebras of $Spin(n,n)$ and $SO(n,n)$ are isomorphic. In addition,
requirement of gauge invariance imposes an extra constraint on the
fluxes that determine the deformations.

\end{narrower}

\vspace{4cm}

\end{titlepage}

\newpage

\tableofcontents

\section{Introduction}\label{introduction}
Double Field Theory (DFT) is a field theory defined on a doubled
space, where the usual coordinates conjugate to momentum modes are
supplemented with dual coordinates that are conjugate to winding
modes \cite{HullZ1,HullZ2,HullZ3,HullZ4}. DFT was originally
constructed on a doubled torus, with the aim of constructing a
manifestly T-duality invariant theory describing the massless
excitations of closed string theory \cite{HullZ1,HullZ2}. Later,
this action was shown to be background independent \cite{HullZ3},
allowing for more general doubled spaces than the doubled torus.
Obviously, the dual coordinates might not have the interpretation
of being conjugate to winding modes on such general spaces.
Construction of DFT builds on earlier work, see
\cite{Tseytlin1,Tseytlin2,Siegel1,Siegel2,Siegel3,Hulleski1,Hulleski2,Hulleski3,Hulleski4}.
For reviews of DFT, see \cite{ZwiebachL, Aldazabal2, HohmLust,
Geissbuhler2}.

On a general doubled space of dimension $2n$, the DFT action has a
manifest $O(n,n)$ symmetry, under which the standard coordinates
combined with the dual ones transform linearly as a vector. The
doubled coordinates must satisfy a set of constraints, called the
weak and the strong constraint and the theory  is consistent only
in those frames in which these constraints are satisfied. It is an
important challenge to relax these constraints, especially the
strong one, as in any such frame   the DFT becomes a rewriting of
standard supergravity, related to it by an $O(n,n)$
transformation. Even in this case, DFT has the virtue of
exhibiting already in ten dimensions (part of) the hidden
symmetries of supergravity, that would only appear  upon
dimensional reduction in its standard formulation. This virtue
should not to be underestimated, as it provides the possibility of
implementing duality twisted reductions of ten dimensional
supergravity  with  duality twists belonging to a larger symmetry
group, that would normally be available only in lower dimensions.

Duality twisted reductions (or generalized Scherk-Schwarz
reductions) are a generalization of Kaluza-Klein reductions, which
introduces into the reduced theory mass terms for various fields,
a non-Abelian gauge symmetry and generates a scalar potential for
the scalar fields \cite{SS}. This is possible if the parent theory
has a global symmetry G, and the reduction anzats for the fields
in the theory is determined according to how they transform under
G. It is natural to study duality twisted reductions of DFT, as it
comes equipped with the large duality group $O(n,n)$, and indeed
this line of work has been pursued by many groups so far
\cite{Grana,Geissbuhler,Aldazabal,parktwist,berman}. In
\cite{Aldazabal,Geissbuhler} it was shown that the duality twisted
reductions of DFT gives in 4 dimensions the electric bosonic
sector of gauged ${\cal{N}} = 4 $ supergravity
\cite{SchonWeidner}. A curious fact which was noted in these works
was that the weak and the strong constraint was never needed to be
imposed on the doubled internal space. This (partial) relaxation
of the strong constraint made the twisted reductions of DFT even
more attractive. Later, in \cite{Grana}, this was made more
explicit, as they showed that the set of conditions to be
satisfied for the consistency of the twisted reduction are in
one-to-one correspondence with the constraints of gauged
supergravity, constituting a weaker set of constraints  compared
to the strong constraint of DFT. Following this, in \cite{Roest},
it was shown that the weakening of the strong constraint in the
twisted reductions of DFT implies that even non-geometric gaugings
of half-maximal supergravity (meaning that they cannot be
T-dualized to gauged supergravities arising from conventional
compactifications of ten-dimensional supergravity) has an uplift
to DFT. Such non-geometric gaugings also arise from
compactifications of string theory with non-geometric flux (see,
for example \cite{Wecht,DallAgata,HullReid}) and the relation of
such compactifications with twisted compactifications of DFT was
explored in various papers, including
\cite{Andriot1,Lust,Andriot,Lee}. We should also note that, the
results of \cite{Grana} was also obtained by \cite{bermanodd}, by
considering the duality twisted reductions of the DFT action they
constructed in terms of a torsionful, flat generalized connection,
called the Weitzenb\"{o}ck connection\footnote{This formulation of
the DFT action has the added advantage that it already includes an
extra term, which has to be added by hand in the original
formulation. This extra term is needed in order to match the 4
dimensional half-maximal gauged supergravity with the theory that
results from the duality twisted reduction of the DFT action.}.

In all of the works cited above, only the reduction of the DFT of
the NS-NS sector of massless  string theory was
studied.\footnote{An exception is the work of \cite{parktwist},
where they also include the reduction of the RR sector. However,
their methods are different from ours, as they perform the twisted
reduction in the semi-covariant formalism of DFT
\cite{semicov1,semicov2}.} The fundamental fields in this sector
are the generalized metric (comprising of the Riemannian metric
and the B-field) and the generalized dilaton. In a frame in which
there is no dependence on the dual coordinates, this sector
becomes the NS-NS sector of string theory. We will hereafter refer
to this frame as the ''supergravity frame''. On the other hand,
the DFT of the RR sector of Type II string theory has also been
constructed by Hohm, Kwak and Zwiebach \cite{dftRR, dftRRkisa} (an
alternative formulation of the RR sector, called the
semi-covariant formulation is given in the papers
\cite{park1,park2}). Likewise, in the supergravity frame, this
action reduces to the action of the democratic formulation of the
RR sector of Type II supergravity. The fundamental fields of this
sector are two $SO(10,10)$-spinor fields, $\ciftS$ and $\chi$. The
latter is a spinor field which encodes the massless p-form fields
of Type II theory. It has to have a fixed chirality, depending on
whether the theory is to describe the DFT of the massless Type IIA
theory or Type IIB theory. The field $\ciftS$ is the spinor
representative of the generalized metric, that is, under the
double covering homomorphism between $Pin(n,n)$ and $O(n,n)$, it
projects to the generalized metric of the NS-NS sector. The action
of this sector has manifest $Spin(10,10)$ invariance (not
$Pin(n,n)$) in order to preserve the fixed chirality of $\chi$.
The action has to be supplemented by a self-duality condition,
which further reduces the duality group to $Spin^+(10,10)$.

The aim of this paper is to study the duality twisted reductions
of the DFT of the RR sector of massless Type II theory, with
twists belonging to the duality group $Spin^+(10,10)$. We study
how the action and the gauge transformation rules reduce and
determine the conditions for the consistency of the reduction and
the closure of the gauge algebra. We also construct the Dirac
operator associated with the $Spin^+(10,10)$ covariant derivative
that arises in the RR sector. In finding the reduced theory, we
find it useful to rewrite the action of \cite{dftRR, dftRRkisa} in
terms of the Mukai pairing, which is a natural bilinear form on
the space of spinors \cite{Chevalley,Mukai,Gualtieri}. The
advantage of this reformulation is that the Mukai pairing is
manifestly $Spin(n,n)$ invariant.  If the duality twist is
introduced via the $Spin^+(10,10)$ element $S$ in the RR sector,
the consistency requires that the NS-NS sector should also be
deformed, via a duality twisted anzats introduced by $U =
\rho(S)$. Here, $\rho$ is the double covering homomorphism between
$Pin(n,n)$ and $O(n,n)$. The fact that Lie algebras of $Spin(n,n)$
and $SO(n,n)$ are isomorphic plays a crucial role in all the
calculations.  We show that the set of conditions required for the
consistency of the reduction of the NS-NS sector are also crucial
for the consistency of the reduction of the RR sector. In
addition, the deformed RR sector is gauge invariant only when the
Dirac operator is nilpotent, which in turn imposes an extra
constraint on the fluxes that determine the deformations. The fact
that such a constraint should arise in the presence of RR fields
has already been noted in \cite{Geissbuhler} and was verified in
\cite{parktwist}.

The plan of the paper is as follows. Section \ref{math} is a
preliminary section on spin representations and the spin group.
Most of the material needed in the calculations for the reduction
is reviewed in this section. In the first part of section
\ref{DFT}, we present a brief review of both sectors of DFT, with
a special emphasis on the RR sector. As the DFT of the RR sector
reduces to the democratic formulation of Type II theory in the
supergravity frame, we start this section by a brief review of the
democratic formulation of Type II supergravity. The rewriting of
the action of \cite{dftRR,dftRRkisa} in terms of the Mukai pairing
is also explained in this section. Section \ref{reductionDFT} is
the main section, where we study the reduction of the action and
the gauge algebra and discuss the conditions for consistency and
closure of the gauge algebra. We finish with a discussion of our
results in section \ref{conc}.

\section{Preliminaries on Spin Representations and The Spin
Group}\label{math}

The purpose of  this preliminary section  is to review the
material, which we will need in the  later sections of the paper.
We closely follow \cite{fulton}.

Let V be an even dimensional (m=2n) real vector space with a
symmetric non-degenerate bilinear form (a metric) $Q$ on it. Then
the orthogonal group $O(V,Q)$ is the space of automorphisms of V
preserving Q : \be O(V,Q) = \{A \in Aut(V) : Q(Av, Aw) = Q(v,w), \
\ \ \forall v,w \in V \} \ee If we restrict this set to the
automorphisms of determinant 1, then we get the subgroup SO(V,Q).
The corresponding orthogonal Lie algebras $so(Q) = o(Q)$ are then
the endomorphisms $A: V \rightarrow V$ such that
 \be
 Q(Av, w) + Q(v, Aw) = 0
 \ee \noindent for al v,w in V. The standard methodology in
 constructing the spin representations of the orthogonal Lie algebra
  is to embed it in the Clifford algebra on V associated to
 the bilinear form Q and use the well-known isomorphisms between the Clifford
 algebras and the matrix algebras.

Given the vector space $V$ and the metric $Q$, one can define the
Clifford algebra $C = Cl(V,Q)$ as the universal algebra which
satisfies the property
 \be \label{cliffordpre}
 \{v, w \} \equiv v . w + w . v = 2 Q(v, w)
 \ee
 Here . is the product on the Clifford algebra. $Cl(V,Q)$ is an associative algebra with unit 1 and as such it
 determines a Lie algebra, with bracket $[a, b] = a . b - b . a $.
 Clifford algebras enjoy nice isomorphisms with various matrix
 algebras (the form of which depends on V and Q) under which the Clifford product
 becomes the matrix multiplication. If $e_1, \cdots, e_m$ form a
 basis of $V$, then the unit element 1 and the products $e_I =
 e_{i_1}. \cdots . e_{i_k}$, for $I = \{i_1 < i_2 < \cdots < i_k
 \}$ form a basis for the $2^m$ dimensional algebra $Cl(V,Q)$. The images of these basis
 elements (of $V$ ) under the isomorphisms with the matrix algebras are
 usually called  $\Gamma$-matrices in the physics literature. The
 Clifford algebra is a $Z_2$ graded algebra and it splits as
 $ C = C^{{\rm even}} \oplus  C^{{\rm odd}}$, where $C^{{\rm even}}
 $ is spanned by products of an even number of elements in $V$ and $
 C^{{\rm odd}}$ is spanned by an odd number of elements of $V$.
 The space $C^{{\rm even}}$ is also a subalgebra and it has half
 the dimension of $C$, that is, it is an algebra of dimension $2^{m-1}$.

 The orthogonal Lie algebra $so(Q)$ embeds  in the even part of the
 Clifford algebra as a Lie subalgebra via the map (for a proof, see \cite{fulton}) $\psi \circ
 \varphi^{-1} : so(Q) \rightarrow C^{{\rm even}}$, where $\psi :
 \wedge^2 V \rightarrow Cl(V,Q)$,
  \be \label{psi}
\psi(a \wedge b) = \frac{1}{2} (a . b - b . a) = a . b - Q(a, b)
\ee and \footnote{Here we
 identify the dual space $V^*$ with $V$ via the bilinear form $Q$ and
 hence $\wedge^2 V \subset End(V) = V \otimes V^* \cong V \otimes
 V.$} \bea \varphi :
 \wedge^2 V & \longrightarrow & so(Q)\subset End(V) \\
      a \wedge b & \longmapsto & \varphi_{a \wedge b} \eea  where $\varphi_{a \wedge b}  $ is given by
\be \label{varphi} \varphi_{a \wedge b} (v) = Q(b, v)a - Q(a, v)b,
\ \ \ a, b \in V. \ee

\noindent Our main interest lies in bilinear forms, which are
non-degenerate and are of signature (n,n). Then a maximally
isotropic subspace is of dimension n. (Recall that a maximally
isotropic subspace of $V$
 is a subspace of maximum possible dimension, on which $Q$ restricts to the
zero-form) Let $W$ be such a subspace and let $W'$ be the
orthogonal complement of $W$ with respect to the bilinear form
$Q$, so that $V = W \oplus W'$. The exterior algebra \be
\wedge^\bullet W = \wedge^0 W \oplus \cdots \oplus \wedge^n W \ee
carries a representation of the Clifford algebra $C$ and hence the
orthogonal Lie algebra $so(Q)$, which is a Lie subalgebra of $C$.
In other words, there exists an isomorphism of algebras between
$C$ and $End(\wedge^\bullet W)$. The ismorphism operates as
follows: for $w \in W$ and $w' \in W'$ one has\footnote{Note that
the usual interior product defined on the subspaces $\wedge^k(V)$
can be extended to the whole exterior algebra by linearity.}
 \be w + w'  \longmapsto l(w)
+ l'(w') \in End(\wedge^\bullet W) \ee where \be
\label{spinorialaction} l(w)\alpha = w \wedge \alpha \ \ \ {\rm
and} \ \ \ l'(w')\alpha = \displaystyle i_{(w')^{\natural}}
\alpha. \ee Here $\alpha \in \wedge^{\bullet}W$ and\footnote{Note
that $Q$ allows one to identify V with the dual space $V^*$ and
under the decomposition $V = W \oplus W'$ the subspace $W$ is
identified with $W'^*$ and $W'$ is identified with $W^*$, hence
$(w')^{\natural} $ is in $W^*$ and contraction with
$(w')^{\natural} $ is well-defined.}  \be \label{contraction}
(w')^{\natural}(w) = 2Q(w, w'). \ee It is straightforward to see
that this defines a representation of the algebra $Cl(V,Q)$ by
verifying that \bea (l(w))^2 = (l'(w'))^2
= 0 \\
\{l(w), l'(w')\} =2 Q(w , w') I. \eea This representation of the
Clifford algebra  carried by $\wedge^\bullet W$ is called the spin
representation. This is an irreducible representation as a
representation of the Clifford algebra, however it is reducible as
a representation of the orthogonal Lie algebra $so(Q) \cong
so(n,n)$, which lies in C. The invariant subspaces of the spin
representation under the action of $so(n,n)$ are denoted by $S^+$
and $S^-$ and corresponds to the decomposition of the exterior
algebra into the sum of even and odd exterior powers. Hence we
have \be S^+ = \wedge^{{\rm even}} W, \ \ \ S^- = \wedge^{{\rm
odd}} W \ee  and \be S = S^+ \oplus S^- \ee where $S =
\wedge^\bullet W $ is the spin representation. The elements of
$S^+$ and $S^-$ are called chiral spinors.

 Inside the Clifford algebra lies an important group, the
group Pin(Q), which in fact turns out to be the double covering
group of $O(Q)$. In order to define it, one needs the following
anti-involution $x \longmapsto x^*$ on the Clifford algebra
determined by \be \label{antiinv} (v_1 \cdot \ldots \cdot  v_k)^*
= (-1)^k v_k \cdot \ldots \cdot v_1 \ee for any $v_1, \ldots, v_k$
in $V$. This is the composite of the main automorphism $\tau : C
\longrightarrow C $ and the main involution $\alpha : C
\longrightarrow C$ determined by \bea \tau(v_1 \cdot \ldots \cdot
v_k) &=& v_k \cdot \ldots \cdot  v_1 \\
\alpha(v_1 \cdot \ldots \cdot  v_k) &=& (-1)^k v_1 \cdot \ldots
\cdot  v_k \eea for  $v_1, \ldots, v_k$ in $V$. Note that $(x +
y)^* = x^* + y^* \ {\rm and} \ (x \cdot y)^* = y^* \cdot x^* $,
which follows from $\tau(x. y) = \tau(y) . \tau(x) \ {\rm and} \
\alpha(x . y) = \alpha(x) . \alpha(y)$.

\noindent Now the group Pin(Q) is defined as a certain subgroup of
the multiplicative group of $C(Q)$: \be Pin(Q) = \{ x \in C(Q) : x
\cdot x^* = \pm 1 \ {\rm and} \ x \cdot V \cdot x^{-1} \subset V
\}. \ee Each element in $Pin(Q)$ determines an endomorphism
$\rho(x)$ of
$V$ by \bea \rho & : & Pin(Q)  \longrightarrow   O(Q) \\
           \label{doublehomom}   \rho(x) & : & \ \ \ \ v \  \ \ \    \longmapsto x \cdot v \cdot
              x^{-1} .
              \eea

\noindent One can show that $\rho$ is a surjective homomorphism,
which preserves the metric $Q$ and its kernel is $\{+ 1, - 1\}$
(for a proof, see \cite{fulton}). If we further demand that $x$
lies in the even part of the Clifford algebra, then the group
becomes the spinor group $Spin(Q)$:

\be Spin(Q) = \{ x \in C(Q)^{{\rm even}} : x \cdot x^* = \pm 1 \
{\rm and} \ x \cdot V \cdot x^* \subset V \},\ee  It is easy to
see that $$ Spin(Q) = Pin (Q) \cap C(Q)^{{\rm even}} = \rho^{-1}
(SO(Q)).$$ Restricting further to the elements in $Spin(Q)$, which
satisfies $x . x^* = + 1$, we obtain the subgroup $Spin^+(Q)$.

The Lie algebra of the group $Spin(Q)$ is a subalgebra of the
Clifford algebra with the usual bracket. It can be shown that this
subalgebra is nothing but the Lie algebra $so(Q)$. In other words,
the derived homomorphism \be \rho' : spin(Q) \longrightarrow so(Q)
 \ee is in fact an isomorphism of the Lie algebras and the right hand side of
 \be \rho'(x)(v) = [x, v],  \ee evaluated in the Clifford algebra (regarding $so(Q)$ and $V$ as subspaces
 of the Clifford algebra)  coincides with the standard action of $so(Q)$ on
 $V$.

\vspace*{0.3cm}

\noindent \textbf{Spinorial Action of $so(n,n)$ and $Spin(n,n)$ on
exterior forms:}  Let us  choose a basis $e_M = \{e^1, \cdots,
e^n, e_1, \cdots, e_n\} = \{e^i, e_i\}$ of $V$ such that \be
Q(e^i, e_j) = \delta^i_{\ j}, \ \ \ Q(e_i, e_j) = Q(e^i, e^j) = 0,
\ \ \forall i,j. \ee With respect to this basis $Q$ is represented
by the matrix $\eta$ \be \eta = \left(\ba{cc}
              0 & I_n \\
              I_n & 0
              \ea\right),
              \ee and the definition
              (\ref{cliffordpre}) becomes
\be \label{clifford} e_M . e_N + e_N . e_M = 2 \eta_{MN}. \ee
Obviously the elements $\{e^1, \cdots, e^n\}$ span an isotropic
subspace $W$ and the elements $\{e_1, \cdots, e_n\}$ span the
orthogonal complement $W'$. The metric $Q$ allows us to identify
$W$ with $W'$, that is, we can raise and lower indices with
$\eta$: $e^M = \eta^{MN} e_N$. Looking back at the maps
(\ref{psi}), (\ref{varphi}), one can calculate that \be e_M \wedge
e_N \in \wedge^2 V \longmapsto T_{MN} \in so(n,n) \ee where the
generators $T_{MN}$ of $so(n,n)$ are endomorphisms of $V$
represented by the antisymmetric matrices \be\label{fund}
  (T_{MN})^{L}{}_{K}  \ = -\eta_{KM}\delta^{L}{}_{N}+ \eta_{KN}\delta^{L}{}_{M} =
  -2\eta_{K[M}\delta_{N]}{}^{L}\;.\ee
Under the isomorphism $\psi \circ \varphi^{-1}$, $T_{MN}$ is
mapped to \be \label{spingen} T_{MN} \longleftrightarrow
\frac{1}{4}(e_M . e_N - e_N . e_M) \equiv \frac{1}{2} e_{MN}. \ee
Note that the standard action of $so(Q)$ on $V$ and its action on
$V$ within the Clifford algebra (when we regard both $so(Q)$ and
$V$ as subspaces of $C$) agree, as it should. That is, we have \be
\label{denk} T_{MN} (e_K) = e_L (T_{MN})^L_{\ K}  = \frac{1}{2}
[e_{MN}, e_K] \ee where the bracket on the right hand side above
is evaluated in the Clifford algebra. Let us note that  we obtain
the more familiar elements $T^{MN}$ by raising the indices of
$T_{MN}$ by $\eta$: \be (T^{PQ})^K_{\ L} = (T_{MN})^K_{\ L}
\eta^{MP} \eta^{NQ} = \eta^{KP} \delta_{L}^{\ Q} - \eta^{KQ}
\delta_L^{\ P} = 2 \eta^{K[P} \delta_L^{\ Q]}. \ee It can be shown
that $T^{MN}$ satisfy the following commutation
 relations:
 \be\label{ODDalgebra}
   \big[ T^{MN},T^{KL}\big] \ = \
   \eta^{MK}\,T^{LN} - \eta^{NK}\,T^{LM}-\eta^{ML}\,T^{KN}+\eta^{NL}\,T^{KM}\;.
 \ee

Now that we know the Clifford algebra elements corresponding to
the generators of the orthogonal Lie algebra, we can immediately
calculate the spinorial action of each generator on forms in the
exterior algebra $\wedge^{\bullet}W$. For this purpose, it is
useful to divide the Lie algebra elements $T_{MN}$ into 3 groups:
$T^{mn}, T_{mn}, T^m_{\ n}$. This corresponds to the decomposition
$\wedge^2 V = \wedge^2(W \oplus W') \cong \wedge^2(W) \oplus
\wedge^2(W') \oplus End(W)$. The spinorial action of these
elements on differential forms can now be easily read off from
(\ref{spinorialaction}):
\bea \label{spinLie1} T^{mn}&:& \alpha \longmapsto \frac{1}{2} e^m \wedge e^n \wedge \alpha, \\
\label{spinLie2} T_{mn}&:& \alpha \longmapsto \frac{1}{2} i_{e_m} i_{e_n} \alpha, \\
\label{spinLie3} T^m_{\ n}&:& \alpha \longmapsto \frac{1}{4}( e^m
\wedge i_{e_n} \alpha -i_{e_n}(e^m \wedge \alpha)) = \frac{1}{2}
(-\delta_n^{\ m}  +e^m \wedge i_{e_n}) \alpha. \eea

\noindent Here, it is important to  note that $i_{e_m} e^n =
2\delta_m^{\ n}$, due to the factor 2 in (\ref{contraction}). It
is more common to work with the basis elements $\psi_M =
\frac{1}{\sqrt{2}} e_M$ which satisfies $\{\psi_m, \psi^n \} =
\delta_m^{\ n}$, so that one has $i_{\psi_m} \psi^n = \delta_m^{\
n}$. Then we have: \bea \label{spinLie4} T^{mn}&:& \alpha \longmapsto    \psi^m \wedge \psi^n \wedge \alpha, \\
\label{spinLie5} T_{mn}&:& \alpha \longmapsto   i_{\psi_m} i_{\psi_n} \alpha, \\
\label{spinLie6} T^m_{\ n}&:& \alpha \longmapsto -\frac{1}{2}
\delta_n^{\ m}  +\psi^m \wedge i_{\psi_n} \alpha. \eea In this
case one should also write the spinor $\alpha \in \wedge^{\bullet}
W$ in terms of the basis elements: $\psi^I =
 \psi^{i_1}. \cdots . \psi^{i_k}$.

By exponentiating the Lie algebra elements $T_{MN}$ in the
fundamental representation, one obtains the identity component
$SO^+(n,n)$ of $SO(n,n)$. A general  group element in the identity
component is of the form $\exp{[\frac{1}{2} \Omega^{MN} T_{MN}]}$.
A simple computation shows that any such element can be written in
terms of  the matrices  given below, where $h_B, h_{\beta}, h_A$
corresponds to the exponentiation of $T^{mn}, T_{mn}, T^m_{\ n}$,
respectively.\footnote{The way we have decomposes the indices
implies that we have \be (T^{MN})^K_{\ L} = \left(\ba{cc}
(T_{MN})_k^{\ l} &
(T_{MN})_{kl} \\
(T_{MN})^{kl} & (T_{MN})^k_{\ l} \ea\right). \ee}
\bea\label{oddelemB} h_B \ &=&
\begin{pmatrix} 1 & -B \\ 0 & 1
\end{pmatrix} \ , \quad B^T = - B \; , \\
\label{oddelembeta} h_{\beta} \
 &=& \begin{pmatrix} 1 &   0 \\ \beta &
1
\end{pmatrix}, \quad  \beta^T = - \beta\\
 \label{oddelemA} h_A \
 &=& \begin{pmatrix} e^A &   0 \\ 0 &
e^{-(A)^T}
\end{pmatrix}  \ ,  \label{PinEl33} \eea Here   we have named $ B_{kl} =
 \Omega_{[kl]}, \beta^{kl} = \Omega^{[kl]}, A^{l}_{\ k}
= \frac{1}{2}(\Omega_k^{\ l} - \Omega^l_{\ k})$. On the other
hand, exponentiation of the generators in the spin representation
gives the corresponding elements $S_B, S_{\beta}, S_A$ in the
identity component $Spin^+(n,n)$ of the spinor group $Spin(n,n)$,
which act on the differential forms as follows: \bea
\label{spinelemB} S_B: \ \ \alpha &\longmapsto & e^{-B} \wedge
\alpha =
(1 - B + \frac{1}{2} B \wedge B - \ldots ) \wedge \alpha, \\
\label{spinelembeta} S_{\beta}: \ \ \alpha &\longmapsto &
e^{\beta} \alpha = (1 + i_{\beta} + \frac{1}{2} i_{\beta}^2 +
\cdots ) \alpha,
\\
\label{spinelemA} S_A: \ \ \alpha &\longmapsto &
\frac{1}{\sqrt{det r}} (e^{A})^* \alpha. \eea  These
transformation rules follow immediately from
(\ref{spinLie1})-(\ref{spinLie3})\footnote{Note that the
determinant term arises from exponentiation of the trace term
which appears in (\ref{spinLie3})}. Here $B = \frac{1}{4} B_{kl}
e^k \wedge e^l = \frac{1}{2} B_{kl} \psi^k \wedge \psi^l, \ \beta
= \frac{1}{4} \beta^{kl} e_k \wedge e_l = \frac{1}{2} \beta^{kl}
\psi_k \wedge \psi_l,$ and $ r = e^{A}$. Also, $i_{\beta} \alpha =
\frac{1}{2} \beta^{ij} i_{\psi_i} (i_{\psi_j} \alpha) $ and $r^*
\alpha = r_j^{\ i} \psi^j \wedge i_{\psi_i} \alpha $, which is the
usual action of $GL^+V$ on forms, where $GL^+V$ is the space of
(orientation preserving) linear transformations on $V$ of strictly
positive determinant. Note that all these elements satisfy $S S^*
= 1$, that is, they lie in the component $Spin^+(n,n)$.

It can be checked that the above elements $h_B, h_{\beta}, h_A$
and the corresponding $S_B, S_{\beta}, S_A$ satisfy $\rho(S) = h$,
by verifying that  (\ref{doublehomom}) is satisfied. In other
words, one can verify that  \be  e_N h^N_{\ M} = S. e_M . S^{-1}.
\ee Multiplying both sides with $\eta^{KM}$ and using the identity
$\eta^{NP} (h^{-1})^K_{\ P} = \eta^{KM} h^N_{\ M}$ we also have
\be \label{oddspin} (h^{-1})^M_{\ N} e^N = S . e^N . S^{-1}.\ee
Note that the right hand-side remains the same if we change $S
\rightarrow -S$, which reflects the fact that the kernel of the
homomorphism $\rho$ is $\{1, -1\}$, that is, $\rho(S) = \rho(-S) =
h$. Obviously, these relations also hold for the Gamma matrices
$\Gamma_M$, which are the matrix images of the Clifford algebra
generators $e_M$ under the isomorphisms with the matrix algebras.
Under such an isomorphism the Clifford multiplication becomes
matrix multiplication and we have\footnote{Here, we abuse the
notation by calling the matrix image of the Clifford algebra
element S also S.} \be \label{mainwithgamma} \Gamma_N h^N_{\ M} =
S \Gamma_M S^{-1}, \ \ \ (h^{-1})^M_{\ N} \Gamma^N = S \Gamma^M
S^{-1}. \ee

Before we move on to the discussion of some important elements of
$Pin(n,n)$, which do not lie in $Spin^+(n,n)$, we would like to
make a remark. Note that the  description of spinors as  forms in
an exterior algebra that we have discussed above is very useful
and one can take this idea one step further by demanding that $W$
is the cotangent space at a point p of an $n$-dimensional smooth
manifold M, $W = T_p^* M$. Then the orthogonal complement is
naturally identified with the tangent space $W'= T_p M$. Then $V =
T_p^* M \oplus T_p M$ is a section of the bundle $T^* \oplus T$.
All the linear algebra discussed above can be transported to the
whole bundle, as it is known that (for example, see
\cite{Gualtieri}) the $SO(n,n)$ bundle $T^* \oplus T$ on an
orientable manifold always carries a $Spin(n,n)$ structure. Then
the spinor fields becomes sections of the exterior bundle
$\wedge^\bullet T^*M$, which are smooth differential forms on $M$,
which are also called polyforms in the physics literature due to
the fact that they are not necessarily homogenous forms. This is
the setting in generalized complex geometry
\cite{Hitchin,Gualtieri}, where the identification of $Spin(n,n)$
spinor fields with smooth differential forms plays a crucial role.

Let us now  move on to the discussion of some other important
elements in $Pin(n,n)$, that will be needed in the rest of this
paper. So far, our aim has been to understand the spinorial action
of the orthogonal Lie algebra (which is isomorphic to the Spinor
Lie algebra) on the exterior algebra $\bigwedge^{\bullet} W$. At
the group level, this has given us only the identity components of
the orthogonal group and the Spinor group. In implementing the
duality twisted reduction, we will only need such elements
(connected to the identity element), as  the real symmetry group
of the RR sector of the DFT action is $Spin^+(n,n)$ (see section
\ref{extension}). However, in constructing this part of the DFT
action, one needs more.  For example,  the spinor representative
of the generalized metric $\cH$ is in $Spin^-(n,n)$, as the
generalized metric itself must be in $SO^-(n,n)$ due to the
Lorentzian signature of the Riemannian metric encoded in $\cH$
\cite{dftRR}. In order to understand such elements, one needs  the
elements of $ O(n,n)$ which interchanges $ e^i \leftrightarrow
e_i$ and keeps all other basis elements of $V$ fixed , possibly up
to a
sign. Let us define the following $O(n,n)$ elements:\be h_i^{\pm} = \pm \left(\ba{cc} 1 - E_i & \pm E_i \\
                                    \pm E_i  & 1 - E_i \ea \right), \ \ (E_i)_{jk} = \delta_{ij}\delta_{ik}. \ee
In the fundamental representation $h_i^+$ ($h_i^-$) interchanges
$e^i \leftrightarrow e_i$ and for all other basis elements it
sends $e^M \rightarrow e^M$ ($e^M \rightarrow -e^M$). One can
easily find the element $\Lambda_i^{\pm}$ in the Pinor group which
projects to these elements. They are given as \be \Lambda_i^{\pm}
= (\psi^i \mp \psi_i),  \ee where we have used the normalized
$\psi^M = \frac{1}{\sqrt{2}} e^M$ so that $\Lambda_i \Lambda_i^* =
\pm 1$, rather than $\pm 2$. Note that $(\Lambda_i^{\pm})^2 = \mp
1$, so we have $(\Lambda_i^+)^{-1} = -\Lambda_i^+$ and
$(\Lambda_i^-)^{-1} = \Lambda_i^-$. One can easily verify the
following by using the Clifford algebra relations \be
 \Lambda_i^{\pm} . e^M . (\Lambda_i^{\pm})^{-1}
                    =\left\{
  \begin{array}{l l l}
   e_i & \quad \text{if\; $e^M = e^i$}\\
   e^i & \quad \text{if\, $e^M = e_i$\,.}\\
   \pm e^M & \quad \text{otherwise}
  \end{array} \right.
                    \ee
Therefore, $\rho(\Lambda_i^{\pm}) = h_i^{\pm}$, as we have
claimed.

From the elements $\Lambda_i^{\pm}$ one can construct a very
important element in the Pinor group, which projects to the
following matrix $J$ in $O(d,d)$: \be \label{j} J = \left(\ba{cc}
0 & I \\
I & 0 \ea \right). \ee Obviously, $J$ swaps $e^i \leftrightarrow
e_i$ \emph{for all} $i$.  On the other hand, $h_i^{\pm} =
\rho(\Lambda_i^{\pm})$ interchanges $e^i $ with $e_i$,  while
keeping all other basis elements fixed, possibly up to a sign $e^M
\leftrightarrow \pm e_M, \ M \neq i$. Therefore, to construct the
Pinor group element that projects to $J$, we need the product
 of all such elements, with some extra care  to determine the overall sign. With a bit
 of work, one can show that   the Pinor group element
$C$, which satisfies $\rho(C) = J$ is \be \label{chargeeven} C =
C^+ \equiv \Lambda_1^+ \ldots \Lambda_d^+, \ee in even dimensions
and \be \label{chargeodd} C = C^- \equiv \Lambda_1^- \ldots
\Lambda_d^- \ee in odd dimensions.

\noindent Note that  $(\Lambda_i^+)^2 = -1$ and $(\Lambda_i^-)^2 =
1$ for all $i$. This implies  that \be C^+ (C^+)^* = 1\ee and \be
C^- (C^-)^* = -1.\ee On the other hand, with a bit of care with
the ordering of the elements one can calculate that \be C^2 =
(-1)^{\sum_1^{d} (d - k) } I = (-1)^{d(d-1)/2} I \ee which gives
\be \label{cters} C^{-1} = (-1)^{d(d-1)/2} C, \ee both for $C^+$
and $C^-$. It is straightforward to check that $C$ indeed
satisfies (both in odd and even dimensions) \be \label{GammaC} C
\, \Gamma^M \, C^{-1} =
  J^M{}_{N} \Gamma^N
\,  \;  \ee (note that $J^{-1} = J$), so indeed \be \rho (C ) \ =
\ J, \, \ee as we have claimed. Since $C$ and $C^{-1}$ just differ
by a sign, we also have $\rho (C^{-1} ) = J$ as a result of which
we have
 \be
\label{GammaCinv} C^{-1} \, \Gamma^M \, C \ = \ J^M{}_{N} \Gamma^N
\; . \ee

\noindent It is appropriate to call this element of the Pinor
group the charge conjugation matrix, as it satisfies the same
Gamma matrix relations as the standard charge conjugation matrix
in quantum field theory. By the help of it, it is possible to
define the action of a dagger operator in the Clifford algebra as
\be \label{dagger-nice} S^\dagger \equiv C\, \tau (S) \,C^{-1} \,.
\ee Obviously, one has $(S_1\cdot S_2)^\dagger = S_2^\dagger \cdot
S_1^\dagger$ (which follows immediately from $\tau(S_1 . S_2) =
\tau(S_2) . \tau(S_1)$) and it can be checked that $ C^\dagger \ =
\ C^{-1} \, $ (as $C \tau(C) = 1$ both in even and odd
dimensions). It is also straightforward to verify that $S \in$
Pin$(n,n)$ implies $S^\dagger \in$ Pin$(n,n)$.  Also note that
$\tau(S) = S^* = \pm S^{-1}$, when $S \in Spin^{\pm}(n,n)$ so we
have \be \label{daggerspin} S^\dagger = C\, S^\star \,C^{-1} = \pm
C\, S^{-1} \,C^{-1} \,, ~~~ S \in \hbox{Spin}^{\pm} (n,n) \,. \ee
The following facts can be proved without much effort (for
details, see \cite{dftRR}) \be \rho(\tau(S)) = \rho(S)^{-1} \ \ \
{\rm and} \ \ \ \rho(S^{\dagger}) = \rho(S)^T. \ee

\noindent \textbf{A bilinear form on the space of spinors: Mukai
pairing:} The last thing we would like to discuss is the natural
inner product on the Clifford module $\wedge^\bullet W$. Later in
section (3.3), we will utilize this inner product in order to
rewrite   the DFT action of the RR sector of Type II theory.
Recall the map $\tau: v_1 \cdot \ldots \cdot v_k \longmapsto  v_k
\cdot \ldots \cdot v_1$ we defined above. It represents a
transpose map in the Clifford algebra which, from the point of
view of the spin module, arises from the following bilinear form
on $\langle \ , \ \rangle : S \otimes S \rightarrow \wedge^n W$:
\be \label{mukai} \langle \chi_1, \chi_2 \rangle = (\tau(\chi_1)
\wedge \chi_2 )_{{\rm top}} = \sum_j (-1)^j (\chi_1^{2j} \wedge
\chi_2^{n-2j} + \chi_1^{2j+1} \wedge \chi_2^{n-2j-1}), \ \ \
\chi_1, \chi_2 \in \wedge^\bullet W, \ee where $( )_{{\rm top}}$
means that  the top degree component of the form should be taken
and the superscript $k$ denotes the $k$-form component of the
form. This bilinear form is known as the Mukai pairing and it
behaves well under the action of the Spin group \cite{Gualtieri}:
\be \label{mukaiinvariant} \langle S \chi_1, S \chi_2 \rangle =
\pm \langle \chi_1 , \chi_2 \rangle, \ \ S \in Spin^{\pm}(n,n).\ee
This bilinear form is non-degenerate and it is symmetric in
dimensions $n \equiv 0, 1 $ (mod 4) and is skew-symmetric
otherwise: \be \label{mukaisymmetric} \langle \chi_1 , \chi_2
\rangle = (-1)^{n(n-1)/2} \langle \chi_2 , \chi_1 \rangle. \ee In
particular, it is skew-symmetric for $n=10$, which is the relevant
dimension in constructing the DFT action for Type II strings. Also
importantly, the bilinear form is zero on $S^+ \times S^-$ and
$S^- \times S^+$ for even $n$ and it is zero on $S^+ \times S^+$
and $S^- \times S^-$ for odd $n$. More details can be found in
\cite{Gualtieri}.

Now assume that there exists an inner product on the vector space
$W$. This also induces a non-degenerate bilinear form  on
$\wedge^\bullet W$ taking values in $\wedge^n W$: \be
\label{hodgeinner} (\chi_1, \chi_2) = \chi_1 \wedge \star \chi_2 =
\sum_j \chi_1^j \wedge \star \chi_2^j \ee where $\star$ is the
Hodge duality operator with respect to the inner product on $W$.
It is possible to show that this bilinear form is related to the
Mukai pairing in the following way: \be \label{mukaihodge}
(\chi_1, \chi_2) = \langle \chi_1, C^{-1} \chi_2 \rangle =
(\tau(\chi_1) \wedge C^{-1} \chi_2 )_{{\rm top}}, \ee where the
charge conjugation matrix presented in (\ref{chargeeven},
\ref{chargeodd}) should be written in terms of an orthonormal
basis with respect to the inner product on $W$.

\section{Democratic Formulation of Type II Theories and the Double Field Theory
Extension}\label{DFT}

\subsection{Democratic Formulation}

The aim of this subsection is to give a brief review of the
democratic formulation of the bosonic sector of (massless) Type
IIA and Type IIB supergravity theories \cite{Fukuma,Bergshoeff}
(also see \cite{Zumino}). The (bosonic) matter
content of these two theories are as follows: \bea IIA : & & \ \ \ \{g, B_2, \phi, C_1, C_3 \} \\
IIB : & & \ \ \ \{g, B_2, \phi, C_0, C_3, C_5 \} \eea The NS-NS
sector, which only involves the metric $g$, the Kalb-Ramond field
$B_2$ (which is a 2-form field) and the dilaton $\phi$ is common
to both Type IIA and Type IIB (as well as to other 3 perturbative
superstring theories) and is given as
 \be \label{NSaction} S_{\rm
NS-NS} =  \int  e^{-2 \phi}  \left[ R + \frac{1}{2} (d \phi \wedge
\star d\phi) - \frac{1}{2} ( H^{(3)} \wedge \star  H^{(3)})
\right] \; , \ee where $H_3 = dB_2$. In order to write down the
Lagrangian for the RR sector in the democratic formulation, one
first defines the following modified RR potentials:
\begin{eqnarray}\label{RR1}
D_0& \equiv & C_0, \ \ \ \ \ \ \ \ \ \ \ \ \ \ \ \ \ \ \ \ \ \ D_1 \equiv C_1, \\
D_2& \equiv & C_2 + B_2 \wedge D_0, \ \ \ \ \ \ \ D_3 \equiv C_3 +
B_2
\wedge C_1, \nonumber \\
D_4& \equiv & C_4 + \frac{1}{2} B_2 \wedge C_2 + \frac{1}{2} B_2
\wedge B_2 \wedge C_0. \nonumber
\end{eqnarray}
Now introduce
\begin{equation}\label{RR2}
D \equiv \sum_{p=0}^8 D_p, \ \ \ \ \ F \equiv e^{-B_2}
\sum_{p=0}^8 dD_p = \sum_{p=0}^8 F_{p+1}.
\end{equation}
The indices run from 0 to 8, as we have also included the
electromagnetic  duals of the gauge potentials $D_p$. The
electromagnetic duals $ D_{8-p}$  of $D_p$ are the potential
fields obtained by solving the field equations for the latter.
This ensures that $F$ defined as above satisfies \be
\label{duality} F_{10-p} = (-1)^{[\frac{p-1}{2}]} * F_p \ee where
$[\frac{p-1}{2}]$ is the first integer greater than or equal to
$\frac{p-1}{2}$. Note that $D$ is a section of the exterior bundle
$\wedge^\bullet T^*M$, where $M$ is the manifold on which the RR
fields live. We can also decompose \be D = D^+ + D^- \ee where
$D^+$ involves k-forms of even degree (k=0,2,4,6,8), whereas $D^-$
involves forms of odd degree. Then $D^+$ and $D^-$ are sections of
the bundles $\wedge^{{\rm even}} T^*M$ and $\wedge^{{\rm
odd}}T^*M$, respectively. Obviously, there is a corresponding
decomposition of the differential form $F = F^+ + F^-$.

\noindent Now consider the following simple actions: \bea
\label{democraticactionIIA} S_{\rm RR}^{IIA} &=& \frac{1}{ 4 }
\int F^+ \wedge * F^+ \  \equiv \ \frac{1}{ 4 } \int \sum_{n = 2,
4, 6, 8}
F^{(n)} \wedge * F^{(n)} \; , \\
\label{democraticactionIIB} S_{\rm RR}^{\rm IIB}
 &=& \ \frac{1}{ 4} \int F^- \wedge * F^- \equiv \  \frac{1}{ 4} \int   \sum_{n = 1, 3, 5, 7, 9}
 F^{(n)} \wedge * F^{(n)} \; . \quad \
\eea

\noindent It can be shown that the actions given above are
equivalent to the standard action  of Type IIA and Type IIB
supergravity theories, which  also involve some complicated
Chern-Simons type terms, in the following
sense\cite{Fukuma,Bergshoeff,dftRR}: If one applies the duality
relations (\ref{duality}) to the field equations derived from the
actions (\ref{democraticactionIIA}), (\ref{democraticactionIIB}),
then one obtains exactly the same field equations that one would
have derived from the standard actions. The field equations for
lower degree form fields match directly in the two formulations.
On the other hand,  the field equations (in the democratic
formulation) for the higher degree fields which are absent in the
standard formulation becomes, after applying (\ref{duality}), the
Bianchi identities for the lower degree fields in the standard
formulation.


\subsection{Double Field Theory Extension}\label{extension}

In the previous section, we have seen that the (modified) RR
fields form sections of the bundles $\wedge^{{\rm odd}} T^*M$ and
$\wedge^{{\rm even}}T^*M$ for Type IIA and Type IIB, respectively.
We have also seen in section 2 that fibers of these bundles, when
$T_p^*M$ is regarded as an isotropic subspace of the doubled
vector space $T_p M \oplus T_p^*M$ at a given point $p \in M$, are
in fact modules for the Clifford algebra $Cl(n,n)$ (when $M$ is
$n$ dimensional) and carry the irreducible spin representation for
the isomorphic Lie algebras $so(n,n), spin(n,n)$ and the
corresponding Lie groups. This structure on the fibers can be
transported to the whole bundle $T \oplus T^*$ on any orientable
manifold $M$. This immediately tells us that the modified RR
fields transform in the spin representation of the  group
$SO(n,n)$ or $Spin(n,n)$. In fact, the main motivation of
constructing the democratic formulation in the first place was to
show the invariance of the RR sector under the orthogonal group
\cite{Fukuma}. In order to achieve this, one reduces Type IIA or
Type IIB on a $(10-d)$-dimensional torus. The invariance of the
scalar and vector fields in $d$ dimensions under $O(d,d)$ had
already been a well-known fact. The vectors transform in the
fundamental representation of $SO(d,d)$, whereas scalar fields
form the coset $SO(d,d) / SO(d) \times SO(d)$ and transform
non-linearly.\footnote{Note that we are restricting ourselves to
$SO(d,d)$ here. In fact the whole Type II theory is invariant
under the bigger group $O(d,d)$, which also involves the T-duality
transformations between the Type IIA and Type IIB theories, given
by the standard Buscher transformation rules in the NS-NS sector.
In the RR sector, this corresponds to changing the chirality of
the spinor state, which is fixed at the outset in the democratic
formulation. Also note that we prefer to keep the dimension $d$
general, rather than fixing it to $d=10$} In \cite{Fukuma}, it was
shown that the RR sector couples to the vector and scalar fields
through the $Spin(d,d)$ matrix, which projects, under the
homomorphism $\rho: Spin(d,d) \rightarrow SO(d,d)$ onto the
$SO(d,d)$ element that parameterizes the $SO(d,d) / SO(d) \times
SO(d)$ scalar coset. They also show that the reduced action can be
put in a form in which the $Spin(d,d)$ invariance is manifest. As
a result, it was established that the $d$ dimensional theory
obtained by dimensional reduction on an ($10-d$)-dimensional torus
was invariant under $SO(d,d)$, not only in the NS-NS sector, but
also in the RR sector.

Double Field Theory (DFT) of Type II strings is an extension  of
massless Type II string theories, in which the duality symmetry
$SO(d,d)$ is already manifest in $d = 10$ dimensions without the
requirement of dimensional reduction.\footnote{In fact, the DFT of
the NS-NS sector of the massless Type II theories is invariant
under the larger group $O(d,d)$. However, the RR sector is only
invariant under $Spin^+(d,d)$.}  The main purpose of this section
is to give a brief overview of DFT and in particular, to review
how the sector of DFT describing the RR fields is an extension of
the democratic formulation of Type II theories, in the sense that
it reduces exactly to it in a particular frame. In what follows,
we will keep the dimension $d$ general, rather than fixing it to
$d=10$, unless it is inevitable.

The main idea in DFT is to allow the (massless) fields in string
theory to depend   on "dual coordinates", in addition to the usual
coordinates of the space-time manifold on which the string
propogates. For backgrounds admitting non-trivial cycles, e.g. for
toroidal backgrounds, the dual coordinates are interpreted as
being conjugate to the winding degrees of freedom, in the same way
space coordinates and momenta are conjugate variables in classical
field theory. This idea in DFT is inspired by closed string field
theory, where all string fields naturally depend on both the usual
coordinates and the dual coordinates. DFT aims to realize this in
the sector of massless fields in order  to construct a manifestly
T-duality invariant action describing this sector. In string
theory, momentum and winding modes combine to transform as a
vector under the T-duality group $O(d,d)$. Therefore, in DFT one
demands  the same behavior from the space-time and dual
coordinates, that is, they form an $O(d,d)$ vector transforming
as: \be X'^M =
h^M_{\ \ N} X^N, \ \ \ \ X^M =  \left(\ba{c} \tilde{x}_i \\
                                              x^i \ea \right) \ee
Here $\tilde{x}_i$ are the  dual coordinates and $h^M_{\ \ N}$ is
a general $O(d,d)$ matrix. In what follows we will always
decompose the indices $M$ labelling the $O(d,d)$ representation as
$^M = (_i, \ ^i)$, where $^i$ and $_i$ label representations of
the $GL(d)$ subgroup of $O(d,d)$. We will raise and lower indices
by the $O(d,d)$ invariant metric $\eta$, so that $X_M = \eta_{MN}
X^N$. Although the theory is formally doubled by the introduction
of the dual coordinates, the existence of an $O(d,d)$ invariant
constraint, called the strong constraint, makes sure that there is
always a choice of a frame in which all the fields and gauge
parameters depend only on  half of the coordinates. The constraint
is $O(d,d)$ invariant and is given below:
  \be\label{ODDconstr}
   \partial^{M}\partial_{M}A \ = \ \eta^{MN}\partial_{M}\partial_{N}A \ = \ 0\;, \qquad
   \partial^{M}A\,\partial_{M}B \ = \ 0\;, \qquad
   \eta^{MN} \ = \  \begin{pmatrix}
    0&1 \\1&0 \end{pmatrix}\,,
 \ee
where $A$ and $B$ represent any fields or parameters of the
theory. To be more precise, the first of the above constraints is
called the \emph{weak constraint} and follows from the level
matching constraint in closed string theory. The second constraint
is stronger and is called the \emph{strong constraint}.  Regarding
the partial derivatives as a coordinate basis for the tangent
space, the strong constraint implies that all vector fields are
sections of a restricted tangent bundle in the sense that at each
point the tangent space is restricted to a maximally isotropic
subspace with respect to the metric $\eta$.

 Let us now present the DFT action, in its generalized metric formulation, which
 was first constructed by Hohm, Hull and Zwiebach for the NS-NS sector
\cite{HullZ4}, and then by Hohm, Kwak and Zwiebach for the RR
sector \cite{dftRR}. These actions can also be presented in terms
of a generalized vielbein, as was first done in \cite{Siegel1}.
\be {\cal{S}} = \int \ dx d\tilde{x} \left( {\cal{L}}_{{\rm
NS-NS}} + {\cal{L}}_{{\rm RR}} \right), \ee where \be
\label{DFTaction1} {\cal{L}}_{{\rm NS-NS}} = e^{-2 d} \ {\cal{R}}
(\cH, d) \ee and
\begin{equation}\label{DFTaction} {\cal{L}}_{{\rm RR}} =  \frac{1}{4}
(\slashed{\partial} \chi)^{\dagger} \ {\mathbb{S}} \
\slashed{\partial} \chi.
\end{equation} This action has to be implemented by the following self-duality constraint
\be \label{selfduality} \slashed{\partial} \chi = - \cK \
\slashed{\partial}\chi, \ \ \ \cK \equiv C^{-1} \ciftS. \ee We
will call the first term in the above action the DFT action of the
NS-NS sector of string theory, whereas the second term will be
referred to as the DFT action of the RR sector. The reason for
this terminology is that in the frame $\tilde{\partial}^i = 0$
(which we call the ''supergravity frame''), which solves the
strong constraint trivially, the first term reduces to the
standard NS-NS action for the massless fields of string theory and
the second term reduces to the RR sector of the democratic
formulation of Type II supergravity theories, discussed in section
(3.1). It is in this sense that this action is an extension of the
democratic formulation of Type II theory.

The term $\cR(\cH, d)$ in (\ref{DFTaction1}) is the generalized
Ricci scalar and its explicit form can be found in \cite{HullZ4}.
It is defined in terms of the generalized metric $\cH$ and the
generalized dilaton $d$. These are $O(d,d)$ covariant tensors (in
fact the dilaton is invariant) depending on both the space-time
and dual coordinates. Their precise form is as below:
\begin{equation}\label{genmetric}
 \cH_{MN} = \left(\begin{array}{cc}
                         \cH^{ij} & \cH^i_{\  j} \\
                         \cH_i^{\  j} & \cH_{ij} \\
                         \end{array}\right) = \left(\begin{array}{cc}
                         g^{ij}  & -g^{ik} b_{kj} \\
                         b_{ik}g^{kj} & g_{ij} - b_{ik} g^{kl} b_{lj}\\
                         \end{array}\right), \ \ \ \ e^{-2 d} =
\sqrt{g} e^{-2 \phi}.
\end{equation} where $\ g = \mid {\rm det} g \mid $. $\cH$ is a symmetric
$O(d,d)$ matrix and as such it satisfies $\cH_{MP} \eta^{PQ}
\cH_{QR} = \eta^{MR}$. 
The   Ramond-Ramond sector couples to the NS-NS sector via
$\ciftS$, where $\ciftS$ is the spinor field which projects to the
generalized metric $\cH$ under the homomorphism $\rho$ of section
2, that is, $\rho(\ciftS) = \cH$. In Lorentzian signature, the
generalized metric $\cH$ is in the coset $SO^-(d,d)$\footnote{When
the space-time metric $g$ is positive definite, so is the
generalized metric $\cH$ and hence its components form a matrix
that lies in $SO^+(d,d)$. In this case the corresponding spin
group element $ {\mathbb{S}}$ is in $Spin^+(d,d)$. However, when
the (semi-)Riemannian metric $g$, has Lorentzian signature, then
$\cH$ is in $SO^-(d,d)$ and correspondingly ${\mathbb{S}}$ lives
in $Spin^-(d,d)$.  Here, $SO^+(d,d)$ is the component of $SO(d,d)$
connected to the identity. It is also a subgroup, whereas its
complement, $SO^-(d,d)$ is  a coset of $SO^+(d,d)$} and there are
subtleties in lifting this to an element $\ciftS$ of $Spin^-(d,d)$
(for a detailed discussion, see \cite{dftRR}). So, in \cite{dftRR}
the following viewpoint was adopted: it is the spin field $\ciftS
\in Spin^-(d,d)$, rather than the generalized metric, which has to
be regarded as the fundamental gravitational field. The
generalized metric $\cH$ is then constructed by projecting onto
the corresponding unique element in $SO^-(d,d)$, so that $\cH =
\rho(\ciftS)$. The field $\ciftS$ satisfies $\ciftS^{\dagger} =
\ciftS$, which immediately implies that $\cH$ is symmetric, as it
has to be.

The other dynamical field in the DFT of the RR sector is the
spinor field $\chi$, which encodes all the (modified) p-form
fields in the RR sector. The field $\chi$ , being a spinor field,
transforms in the spinor representation of $Spin(d,d)$. Its
chirality has to be fixed at the outset, so that it is either an
element of $S^+$ or $S^-$ (see section 2). If we demand that the
doubled manifold $M^{\rm doub}$ is spin and the physical manifold
$M$ sits in it in such a way that at each point $p \in M$, the
cotangent space $T^*_p M$  is an isotropic subspace of the whole
cotangent space $T^*_p M^{\rm doub}$ with respect to the metric
$\eta$, then $\chi$ forms a section of the exterior bundle
$\wedge^{{\rm even}} T^*M$ or $\wedge^{{\rm odd}} T^*M$, depending
on its fixed chirality. Therefore, when restricted to the physical
manifold, that is, in the frame $\tilde{\partial}^i = 0$, it
encodes all the RR fields of  either the Type IIA or the Type IIB
theory, depending on how its chirality has been fixed. More
generally, all the independent fields, including $\chi$, might
depend both on the physical coordinates and the dual ones. The
operator $\slashed{\partial}$ in the action (\ref{DFTaction}),
which differentiates $\chi$ is the generalized Dirac operator
defined as\footnote{From this section on, we will always work with
the Gamma matrices $\Gamma_M$, which are the matrix images of the
Clifford algebra generators $e_M$.} \be \label{diracop}
\slashed{\partial} \equiv \frac{1}{\sqrt{2}} \Gamma^M
\partial_M = \frac{1}{\sqrt{2}}(\Gamma^i \partial_i + \Gamma_i
\tilde{\partial}^i). \ee The self-duality constraint
(\ref{selfduality}) makes sure that the p-form fields encoded by
the spinor field $\chi$ obey the self-duality relations in the
previous section . It should be noted that (\ref{selfduality}) is
consistent only if $\cK^2 = 1$. On the other hand, \be
\label{kkare} \cK^2 = C^{-1} \ciftS C^{-1} \ciftS = C \ciftS C
\ciftS = - C^2 = -(-1)^{d(d-1)/2}, \ee where we have used
(\ref{daggerspin}), (\ref{cters}) and the facts that $ \ciftS \in
Spin^-(n,n)$ and $\ciftS^{\dagger} = \ciftS $. As a result,
consistency of the self-duality equation imposes that $d(d-1)/2$
should be odd, that is $d \equiv 2,3$ (mod 4). These are exactly
the dimensions for which the Mukai pairing is anti-symmetric. This
fact will play a crucial role in section (3.3).

An important ingredient in DFT is the generalized Lie derivative
$\hat{\cL}$, which determines the gauge transformations of the DFT
and   the C-bracket, which determines how the gauge algebra closes
\cite{HullZ2}. Let us define $\xi^M = (\tilde{\xi}_i,\xi^i)$ as
the $O(d,d)$ vector which generates the following gauge
transformations. \bea\label{manifestH}
  \delta_{\xi}{\cal H}_{MN} \ &=& \  \widehat{\cal L}_{\xi} {\cal H}_{MN} \\
  & \equiv & \ \xi^{P}\partial_{P}{\cal H}_{MN}
  +\big(\partial_{M}\xi^{P} -\partial^{P}\xi_{M}\big)\,{\cal H}_{PN}
  +
  \big(\partial_{N}\xi^{P} -\partial^{P}\xi_{N}\big)\,{\cal H}_{MP}\;, \nonumber \\
  \delta d ~\ &=& \ \xi^M \partial_M d - {1\over 2}  \partial_M \xi^M
  \, \nonumber
 \eea in the NS-NS sector and
\bea \label{gaugechi} ~
 \delta_\xi \chi \ = \ \widehat{\cal L}_\xi \chi  \ & \equiv & \ \xi^M \partial_M  \chi
\ + \   {1\over \sqrt{2}}\, \slashed{\partial} \xi^M
\Gamma_{\hskip-1pt M}\, \chi  \, \\
&= & \ \xi^M \partial_M  \chi  \ + \ {1\over 2} \,  \partial_N
\xi_M \Gamma^N \Gamma^M \chi\;. ~~ \nonumber \eea \be
\label{gaugeK} \delta_{\xi} {\cal K}  \ = \ \xi^M
\partial_M  {\cal K} + {1 \over 2} \big[ \Gamma^{PQ}  , \,  {\cal
K} \, \big]  \partial_{P} \xi_{Q} \; , \ee in the RR sector, where
$\Gamma^{PQ} \equiv \frac{1}{2}  [\Gamma^P , \Gamma^Q ]$, as in
(\ref{spingen}).

It was shown in \cite{HullZ2,HullZ4}(for the NS-NS sector) and in
\cite{dftRR}(in the RR sector) that the DFT action is invariant
under these gauge transformations. The gauge transformations in
the RR sector were determined by demanding that they leave the
action invariant as well as demanding compatibility with the gauge
transformation rules in the NS-NS sector.

In the frame $\tilde{\partial}^i = 0$, the gauge parameter
$\xi^{M}=(\tilde{\xi}_{i},\xi^{i})$  combines the diffeomorphism
parameter $\xi^{i}(x)$ and the Kalb-Ramond gauge parameter
$\tilde{\xi}_{i}(x)$. The double field theory version of the
abelian gauge symmetry of p-form gauge fields is \be
\label{gaugelambda} \delta_{\lambda} \chi = \slashed\del \lambda =
\frac{1}{\sqrt{2}} \Gamma^M \partial_M \lambda,\ee where $\lambda
$ is a space-time dependent spinor.

These gauge transformations form a gauge algebra with respect to
the C-bracket, which is the $O(d,d)$ covariantization of the
Courant bracket in generalized geometry \cite{Hitchin,Gualtieri}.
The C-bracket of two $O(d,d)$ vectors is given as \be
\label{cbracket} \big[ \xi_1, \ \xi_2 \big]_C^M = 2 \xi_{[ 1}^N
\del_N \xi_{2 ]}^M - \xi_{[ 1}^P \del^M \xi_{2 ]  P} \ee The gauge
transformations above satisfy \bea \label{closure}
\big[ \delta_{\xi_1}, \delta_{\xi_2}\big] &=& -\delta_{[\xi_1, \xi_2 ]_C} \nonumber \\
\big[ \delta_{\lambda}, \delta_{\xi} \big] &=&
\delta_{\widehat{\cal L}_{\xi}}  \eea We would like to emphasize
that the strong constraint is crucial in proving the closure of
the gauge algebra.

\noindent The DFT action presented in (\ref{DFTaction}) is
invariant under the following transformations: \be
\label{transform}
 \ciftS(X) ~~  \longrightarrow \ciftS^{\prime}(X^{\prime}) \ = \ (S^{-1})^{\dagger}\, \ciftS(X)
 \,S^{-1}\;, ~~ \chi(X) \longrightarrow \chi(X') = S \chi(X)
 \ee
Here $S \in Spin^+(d,d)$  and $X^{\prime}= hX$,  where $h =
\rho(S) \in SO(d,d)$. The dilaton is invariant. The duality group
is broken to $Spin(d,d)$ as the full $Pin(d,d)$ does not preserve
the fixed chirality of the spinor field $\chi$. Also, a general
$Spin(d,d)$ transformation does not preserve the self-duality
constraint (\ref{selfduality}) and the  duality group is further
reduced to the subgroup $Spin^+(d,d)$.\footnote{ The
transformation of $\chi$ in (\ref{transform}) implies that
$\slashed{\partial}\chi \rightarrow S \slashed{\partial} \chi$ and
we have $C^{-1} (S^{-1})^{\dagger} = S C^{-1}$ only for $S \in
Spin^+(d,d)$.}The transformation of $\ciftS$ implies the following
transformation rule for the generalized metric $\cH =
\rho(\ciftS)$: \be \label{transform2}
 \cH(X) ~~  \longrightarrow \cH^{\prime}(X^{\prime}) \ = \ (h^{-1})^{T}\, \cH(X)
 \,h^{-1}\;.
 \ee
These transformation rules will dictate our duality twisted
reduction anzats in  section \ref{reductionDFT}.

\subsection{The DFT Action of the RR Sector Rewritten With the
Mukai Pairing}

In this section, we rewrite the DFT action of the RR sector in
terms of the Mukai pairing reviewed in Section 2. Writing the
action in this form will simplify the calculations, when we study
the duality twisted reduction of the action. Besides, the fact
that the DFT action (\ref{DFTaction}) is an extension of the
democratic formulation of supergravity theory becomes explicit in
this reformulation.

 Recall that the DFT action (\ref{DFTaction}), which was constructed in \cite{dftRR} reduces to
(\ref{democraticactionIIA}) or (\ref{democraticactionIIB}) in the
supergravity frame $\tilde{\partial}^i = 0$, depending on the
chirality of $\chi$. Here, we will start with the supergravity
actions (\ref{democraticactionIIA}) or (\ref{democraticactionIIB})
and show that they extend to the action (\ref{DFTaction}),
rewritten with the Mukai pairing.

The actions (\ref{democraticactionIIA}) or
(\ref{democraticactionIIB}) are quite simple; in fact they just
involve the inner product of $F^{\pm} \in S^{\pm}$ with itself,
where the inner product is the natural inner product
(\ref{hodgeinner}). In section 2, we stated how this inner product
is related with the Mukai pairing, see (\ref{mukaihodge}).
Therefore these Lagrangians can also be written as \be L_{IIA} =
\frac{1}{4} \langle F^{+} , C^{-1} F^{+} \rangle, \ee and \be
L_{IIB} = \frac{1}{4} \langle F^{-} , C^{-1} F^{-} \rangle, \ee
where $\langle , \rangle$ is the Mukai pairing in (\ref{mukai}).
As a matter of fact, we could just as well write \be L =
\frac{1}{4} \langle F, C^{-1} F \rangle \ee with $F = F^+ + F^-$,
as the Mukai pairing is already zero on $S^+ \times S^-$ and $S^-
\times S^+$ for even $d$ and is zero on $S^+ \times S^+$ and $S^-
\times S^-$ for odd $d$.\footnote{Note that $C^{-1} F^{\pm} \in
S^{\pm}$ in even dimensions and $C^{-1} F^{\pm} \in S^{\mp}$ in
odd dimensions.} Hence, there is no need to fix the chirality in
this case; the Mukai pairing already picks up the desired
combinations. Recall that the charge conjugation matrix has to be
written in terms of an orthonormal basis with respect to the
metric on $M$. Alternatively, we can write $C$  as in
(\ref{chargeeven},\ref{chargeodd}) and compensate that by pulling
back the differential form $F$ with the spin representative
$S_g^{-1}$ of the inverse metric $g^{-1}$.\footnote{Note that, for
Riemannian $g$, this operator is just $S_g = S_e S_e^\dagger$,
where $g = e e^t$ and $S_e$ is as in (\ref{spinelemA}) with $A =
e$. For Lorentzian metric, it is a bit more involved, for details
see \cite{dftRR}. For our purposes, it is sufficient to know that
$S_g \in Spin^-(10,10)$ and it satisfies $S_{g^{-1}} = S_g^{-1}$
and $S_g = S_g^\dagger$.} This gives us \be \label{mukaiaction1} L
= \frac{1}{4} \langle F, \ C^{-1} S_g^{-1} F \rangle. \ee Now, it
follows from (\ref{RR2}) that $F = S_b \ \slashed{\partial} \chi$,
where $S_b$ is as in (\ref{spinelemB}) and $\chi$ is the spinor
field encoding the modified gauge potentials $D_p$, see
(\ref{RR1}),(\ref{RR2}). Writing (\ref{mukaiaction1}) in terms of
$\chi$ we have \be \label{mukaiaction2} L = \frac{1}{4} \langle
S_b \ \slashed{\partial} \chi, \ C^{-1} S_g^{-1} S_b \
\slashed{\partial} \chi \rangle. \ee Now we use the invariance
property (\ref{mukaiinvariant}), which gives \be
\label{mukaiaction3} L = \frac{1}{4} \langle \ \slashed{\partial}
\chi, \ S_b^{-1} C^{-1} S_g^{-1} S_b \ \slashed{\partial} \chi
\rangle. \ee Note that the + sign has to be picked in
(\ref{mukaiinvariant}) as $S_b \in Spin^+(10,10)$, as discussed in
section \ref{math}. Now we use (\ref{daggerspin}) to write this
Lagrangian as \be \label{mukaiaction4} L = \frac{1}{4} \langle  \
\slashed{\partial} \chi, \ C^{-1} S_b^\dagger S_g^{-1} S_b \
\slashed{\partial} \chi \rangle. \ee The expression $S_b^\dagger
S_g^{-1} S_b$ that appears above is nothing but the definition of
$\ciftS$ in \cite{dftRR}, so our action becomes \be
\label{mukaiactionfinal} L = \frac{1}{4} \langle  \
\slashed{\partial} \chi, \ C^{-1} \ciftS \ \slashed{\partial} \chi
\rangle. \ee When $\chi = \chi(x)$ and $\ciftS = \ciftS(x)$, this
action is just a rewriting of the supergravity actions
(\ref{democraticactionIIA}) and (\ref{democraticactionIIB}) in the
democratic formulation. On the other hand, when $\chi = \chi(x,
\tilde{x})$ and $\ciftS = \ciftS(x, \tilde{x})$,  the action
(\ref{mukaiactionfinal}) is equivalent to (\ref{DFTaction}) of
\cite{dftRR,dftRRkisa}. Note that, in the first case we have
$\slashed{\partial} \chi(x) = \psi^i
\partial_i \chi(x)$, whereas in the DFT extension we have
$\slashed{\partial} \chi(x, \tilde{x}) = \psi^i \partial_i \chi(x,
\tilde{x}) + \psi_i \tilde{\partial}^i \chi(x, \tilde{x})$.

Let us discuss the transformation properties of this action under
(\ref{transform}). First of all, note that under $ \chi
\rightarrow S \chi$ we have $\slashed{\partial} \chi \rightarrow S
\slashed{\partial} \chi$. Indeed, \bea \slashed{\partial} \chi
\rightarrow  \psi^M (h^{-1})^N_{\ M}
\partial_N (S \chi) & = & S S^{-1} \psi^M S (h^{-1})^N_{\ M}
\partial_N  \chi \nonumber \\
& = & S h^M_{\ P} \psi^P (h^{-1})^N_{\ M}
\partial_N  \chi = S \psi^P \partial_P \chi = S \slashed{\partial} \chi. \eea Here  $ h = \rho(S)$ and we have used (\ref{mainwithgamma})
(recall that $\psi^M = 1/\sqrt{2} \Gamma^M$). As a result, under
(\ref{transform}), the Lagrangian (\ref{mukaiactionfinal})
transforms as \be L \rightarrow \langle S \slashed{\partial} \chi,
C^{-1} (S^{-1})^\dagger \ciftS S^{-1} S \slashed{\partial} \chi
\rangle =  \langle S \slashed{\partial} \chi,  \pm  S \ C^{-1} \
\ciftS \slashed{\partial} \chi \rangle, \ \ \ S \in
Spin^{\pm}(10,10), \ee where we have used (\ref{daggerspin}). Now
the invariance property (\ref{mukaiinvariant}) of the Mukai
pairing immediately implies that the Lagrangian is invariant under
the \emph{whole} $Spin(10,10)$. As we noted above,
 the democratic action (without introducing the dual
coordinates) is already in the form (\ref{mukaiactionfinal}).
However, this action is not invariant under $Spin(10,10)$ unless
we introduce the dual coordinates. Indeed, as can be seen from our
discussion above, $\chi \rightarrow S \chi$ implies
$\slashed{\partial} \chi \rightarrow S \slashed{\partial} \chi$
only when the dual coordinates are introduced.

Recall that the self-duality relation (\ref{selfduality}) involved
the spin element $\cK \in Pin(d,d)$, which we defined as $\cK =
C^{-1} \ciftS$. Consistency imposed $\cK^2 = 1$, which implied
that $d$ has to satisfy $d \equiv 2,3$ (mod 4), since $\cK^2 =
-(-1)^{d(d-1)/2}$, see (\ref{kkare}). It is  possible to rewrite
(\ref{mukaiactionfinal}) as \be \label{mukaiactionK} L   =
\frac{1}{4} \langle \slashed{\partial} \chi,  \ \cK \
\slashed{\partial} \chi \rangle. \ee Note that for even $d$, $\cK
\in Spin^-(d,d)$. Using the invariance property
(\ref{mukaiinvariant}) we then have (for even $d$) \be L = -
\frac{1}{4}  \langle \cK \ \slashed{\partial} \chi, \ \cK^2 \
\slashed{\partial} \chi \rangle. \ee  Now we use (\ref{kkare}) to
write  \be L = (-1)^{d(d-1)/2} \frac{1}{4} \langle \cK \
\slashed{\partial} \chi, \ \slashed{\partial} \chi \rangle \ee It
is an important consistency check that the right hand side above
can be written as \be  \frac{1}{4} \langle \slashed{\partial}
\chi, \ \cK \ \slashed{\partial} \chi \rangle, \ee which follows
immediately from (\ref{mukaisymmetric}).

 When we impose the constraint (\ref{selfduality}) in the
action (\ref{mukaiactionK}), we get \be L = -\frac{1}{4} \langle
\slashed{\partial} \chi, \slashed{\partial} \chi \rangle, \ee
which becomes identically zero for $d \equiv 2,3$ (mod4) due to
the antisymmetry property of the Mukai pairing in these
dimensions. These are exactly the dimensions in which it is
consistent to impose the constraint (\ref{selfduality}). This is
the usual case with constrained actions and as usual, one must
impose the constraint only to the equations of motion, not the
action itself.


\section{Duality Twisted Reductions of DFT: Gauged Double Field Theory}\label{reductionDFT}

In the previous section we reviewed the action of DFT describing
both the NS-NS and R-R sectors of massless string theory. The DFT
action of the NS-NS sector has  global $Pin(d,d)$ symmetry. When
one includes the RR sector, this symmetry group is reduced to
$Spin(d,d)$ due to the chirality condition and is further reduced
to $Spin^+(d,d)$ due to the existence of the self-duality
constraint (\ref{selfduality}). This global symmetry group makes
it possible to implement a duality twisted anzats in the
dimensional reduction of the DFT action. More precisely, the
transformation rule (\ref{transform}) for the fundamental fields
in the theory make it possible to introduce the following duality
twisted dimensional reduction anzats: \bea \label{anzats}
\ciftS(X, Y) &=& (S^{-1})^{\dag}(Y) \ciftS(X) S^{-1}(Y) \\
\label{anzatschi} \chi(X, Y) &=& S(Y) \chi(X) \eea Here, $X$
denote collectively the coordinates of the reduced theory, whereas
$Y$ denote the internal coordinates, which are to be integrated
out. The twist matrix $S(Y)$ belongs to the duality group
$Spin^+(d,d)$ and encodes the whole dependence on the internal
coordinates.

The above anzats for the spinor fields implies the following
anzats in the NS-NS sector: \be \label{anzats2} \cH_{MN}(X, Y) =
U^A_{\ M}(Y) \cH_{AB}(X) U^B_{\ N}(Y).
 \ee The duality twisted dimensional reduction of the DFT action of the
NS-NS sector with the anzats (\ref{anzats2})  has already been
studied by several groups \cite{Geissbuhler,Aldazabal,Grana}, and
the resulting theory was dubbed Gauged Double Field Theory (GDFT)
\cite{Grana}. For the details of the reduction of the action and
the gauge transformations of the dimensionally reduced theory, we
refer the reader to these papers. Here, we also study the duality
twisted reduction of the DFT action describing the RR sector.

In the reduction of the NS-NS sector, it is also possible to
introduce  the following anzats for the generalized dilaton
\cite{Grana} \be \label{anzatsdilaton} d(X, Y) = d(X) + \rho(Y).
\ee This then leads to an overall conformal rescaling in the NS-NS
sector \be \cL_{NS-NS} \rightarrow e^{-2\rho(Y)} \cL_{NS-NS}. \ee
This overall factor contributes to the volume factor, when one
integrates out the $Y$ coordinates in order to define  the GDFT
action of the NS-NS sector \cite{Grana}: \be S_{GDFT} = v \int d^N
X e^{-2d} (\cR + \cR_f) \ee where $\cR_f$ is determined by the
fluxes $f_{ABC}$ and $\eta_A$, as we will discuss in the next
subsection and $v$ is defined as \be \label{v} v = \int d^d Y
e^{-2\rho(Y)}. \ee In the presence of the RR fields, the GDFT
action will be of the form \be S_{GDFT} = v \int d^N X
\left[\left( {\cal{L}}_{{\rm NS-NS}} + {\cal{L}}_{{\rm RR}}
\right) + \left( {\cal{L}}_{{\rm NS-NS}} + {\cal{L}}_{{\rm RR}}
\right)_{{\rm def}}\right] \ee In order to induce the overall
$\rho$-dependent factor in the RR sector, it is necessary to
modify the anzats (\ref{anzatschi}) as follows \be
\label{anzatschi2} \chi(X, Y) = e^{-\rho(Y)} S(Y) \chi(X). \ee

In the next two subsections, we will study the GDFT action arising
from the introduction of the anzatse (\ref{anzats},\ref{anzats2},
\ref{anzatsdilaton}) and (\ref{anzatschi2}). Before we move on,
let us clarify a point. Comparing (\ref{anzats2}) with
(\ref{transform2}), we see that we have $U = h^{-1}$, where $h =
\rho(S) \in SO^+(d,d)$. In
 other words we have $ U = (\rho(S))^{-1} = \rho(S^{-1}). $ The reason that we have made  this naming
 (rather than naming $\rho(S) = U$) is to make sure that our notation is consistent with that of the papers mentioned above, especially that of \cite{Grana}.
Again, following \cite{Grana}, we make a distinction between the
indices of the parent theory and the  indices of the resulting
theory, which we label by $M$ and $A$, respectively.

\subsection{Review of the Reduction of the NS-NS
Sector}

In the duality twisted reduction of the NS-NS sector, there are
two main conditions to be imposed on the twist matrix $U$:
Firstly, one demands that the Lorentzian coordinates $X$ remain
untwisted, which is ensured if the following condition is
satisfied by all the $X$ dependent fields of the resulting GDFT:
\be \label{cond1}
 (U^{-1})^M_{\ A} \partial_M g(X) = \partial_A g(X). \ee
 The second condition is
\be \label{cond2}
 \partial^P (U^{-1})^M_{\ A} \partial_P g(X) = 0. \ee
 This is trivially
satisfied if one works with twist matrices such that  a given
coordinate and its dual are either both external or both internal.
If the anzats involves a non-zero $\rho(Y)$ in
(\ref{anzatsdilaton}), a condition similar to (\ref{cond2}) has to
be imposed also on $\rho$: \be \label{condlambda} \partial^P \rho
\partial_P g(X) = 0.\ee  As was shown in
\cite{Geissbuhler,Aldazabal,Grana}, all the information about the
twist matrix $U$ is encoded in the entities $f_{ABC}$ and $\eta_A$
that we will define below. These entities, which we will refer to
as ''fluxes'', as is usual in the literature, determine both the
deformation of the action and that of the gauge algebra. The
situation is entirely the same in the RR sector as we will discuss
shortly. The fluxes are defined as \be \label{structure} f_{ABC} =
3 \Omega_{[ABC]}, \ \ \ \eta_A =
\partial_M  (U^{-1})^M_{\ A} - 2(U^{-1})^M_{\ A} \partial_M \rho \ee where $\rho$ is as in (\ref{anzatsdilaton}) and \be \label{omega}
\Omega_{ABC} = -(U^{-1})^M_{\ A} \del_M (U^{-1})^N_{\ B} U^D_{\
N}\eta_{CD}.  \ee Note that $\Omega_{ABC}$ are antisymmetric in
the last two indices: $\Omega_{ABC} = - \Omega_{ACB}$. We also
make the following definition \be \label{fa} f_A = -\partial_M
(U^{-1})^M_{\ A}  = \Omega^C_{\ AC} \ee

 It can be shown that  the conditions (\ref{cond1}) and
(\ref{cond2}) imply that the following has to be satisfied: \be
\label{onemlikosul} f^A_{\ BC} \del_A g(X) = 0, \ \ f^A \del_A
g(X) = 0. \ee Note that the second condition in
(\ref{onemlikosul}) and (\ref{condlambda}) imply together that
$\eta_A$ should also satisfy \be \label{onemlikosul2} \eta^A
\del_A g(X) = 0. \ee

These constraints are crucial for the closure of the gauge
algebra. In addition, one also needs that all the fluxes $f_{ABC}$
and $\eta_A$ must be constant. This ensures that the $Y$
dependence in the GDFT is completely integrated out. Also, the
weak and the strong constraint has to be imposed on the external
space so that \be \label{external}
\partial_A \partial^A V(X) = 0, \ \
\partial_A V(X) \partial^A W(X) = 0 \ee for any fields or gauge
parameters $V, W$ that has dependence on the coordinates of the
external space only. Finally, the following Jacobi identity and
the orthogonality condition should
 be satisfied for the closure of the gauge algebra: \be \label{jacobi} f_{E[AB} f_{C]D}^{\ \ \
E} = 0,\ee and\footnote{This condition does not appear in
\cite{Grana}, as they constrain $\eta_A = 0$.} \be
\label{orthogonality} \eta^A f_{ABC} = 0.\ee

To summarize, for the consistency of the reduction of the DFT of
the NS-NS sector one needs the conditions
(\ref{cond1}-\ref{condlambda}) and
(\ref{onemlikosul}-\ref{orthogonality}). In addition, the fluxes
$f_{ABC}$ and $\eta_A$  must be constant. These are the only
conditions that have to be satisfied in order to obtain a
consistent GDFT.

Surprisingly, it is not necessary to impose the strong constraint
in the internal space, that is, one does not need to impose \be
\partial^P U^A_{\ M}
\partial_P U^B_{\ N}. \ee Therefore, the duality twisted anzats
(\ref{anzats})-(\ref{anzatsdilaton}) allows for a relaxation of
the strong constraint on the total space.

\subsection{Reduction of The RR sector}

\noindent Our aim here is to study the reduction of
(\ref{mukaiactionfinal}) and the constraint (\ref{selfduality}).
Recall the main relation (\ref{mainwithgamma}), which we rewrite
here for $U = \rho(S^{-1})$:  \be \label{esas} S^{-1} \Gamma^M S =
(U^{-1})^M_{\ A} \Gamma^A \ee Now, we plug in the anzats \be
\chi(X,Y) = e^{-\rho(Y)} S(Y) \chi(X) \ee  in
$\slashed{\del}\chi(X,Y)$ in (\ref{mukaiactionfinal}) to get \bea
\sqrt{2} \ \slashed{\del}\chi(X,Y) &=& \Gamma^M \del_M \chi(X,Y) =
\Gamma^M \del_M (e^{-\rho(Y)} S(Y) \chi(X)) \nonumber \\
&=& e^{-\rho(Y)} \left\{- \Gamma^M S \ \del_M \rho(Y)  + \Gamma^M S \ \del_M  + \Gamma^M S (S^{-1} \partial_M S)\right\} \chi(X) \nonumber \\
&=&  e^{-\rho(Y)} S(Y) \Gamma^A  \left(-(U^{-1})^M_{\ A} \del_M
\rho(Y) +
 \del_A + (U^{-1})^M_{\ A} S^{-1} \del_M S \right)
\chi(X), \eea where, in passing from the second line to the third,
we have used (\ref{cond1}) and (\ref{esas}).

 Recall that  the Lie algebras of $Spin(d,d)$ and
$O(d,d)$ are isomorphic.
 This gives us the important  property:
\be \label{isom}  \Gamma^A \ (U^{-1})^M_{ \ \ A}  S^{-1}
\partial_M \ S = \frac{1}{4} \Omega_{ABC} \Gamma^A \ \Gamma^B \
\Gamma^C, \ee which we prove now.

 As $U$ and $S$ are in the connected component of the
orthogonal group and the Spinor group, they can be written as \be
  \bigl[ \exp\big(\tfrac{1}{2}\Lambda_{PQ}(Y)T^{PQ}\big)\bigr],
 \ee
where the generators $T_{MN}$ are in the fundamental
representation for the $SO^+(D,D)$ matrix $U$, whereas it is in
the spinor representation for $S$, see section \ref{math}.
Therefore, we have \be
   (U^{-1})^{M}{}_{A} \ = \ \bigl[
   \exp\big(\tfrac{1}{2}\Lambda_{PQ}(Y)T^{PQ}\big)\bigr]^{M}{}_{A}\;,\quad
   \text{and} \quad S = \bigl[\exp\big(\tfrac{1}{2}\Lambda_{PQ}(Y)\frac{1}{2}\Gamma^{PQ}\big)\bigr]
 \ee

\noindent Now we prove (\ref{isom}) starting from the right hand
side:\bea \frac{1}{4} \Omega_{ABC} \Gamma^B \Gamma^C & = &
-\frac{1}{4} (U^{-1})^M_{\ A} \big(U^D_{\ N} \del_M (U^{-1})^N_{\
B}\big) \eta_{CD}
\Gamma^B \Gamma^C \nonumber \\
& = & -\frac{1}{4} (U^{-1})^M_{\ A} \big(U \del_M U^{-1}
\big)^D_{\ B} \eta_{CD}
\Gamma^B \Gamma^C \nonumber \\
& = & -\frac{1}{4} (U^{-1})^M_{\ A} \frac{1}{2} \del_M
\Lambda_{PQ}
\big(T^{PQ}\big)^D_{\ B} \eta_{CD} \Gamma^B \Gamma^C \nonumber \\
& =& -\frac{1}{4} (U^{-1})^M_{\ A} \frac{1}{2} \del_M \Lambda_{PQ}
(\eta^{DP} \delta^Q_{\ B} - \eta^{DQ} \delta^P_{\ B}) \eta_{CD} \Gamma^B \Gamma^C \nonumber \\
& = & -\frac{1}{4} (U^{-1})^M_{\ A} \frac{1}{2} (\del_M
\Lambda_{CB} - \del_M \Lambda_{BC}) \Gamma^B \Gamma^C \nonumber \\
& = & \frac{1}{2} (U^{-1})^M_{\ A}  \del_M \Lambda_{BC}
\frac{1}{4} (\Gamma^B \Gamma^C - \Gamma^C \Gamma^B) \nonumber \\
& = & (U^{-1})^M_{\ A} S^{-1} \del_M S \eea which immediately
implies (\ref{isom}). As a result, we have: \be \label{redchi}
\slashed{\del}\chi(X,Y) = \frac{1}{\sqrt{2}} \Gamma^M \del_M
\chi(X,Y) = e^{-\rho(Y)} S(Y) \slashed{\nabla} \chi(X),
  \ee  where we have
defined \be \label{hatchi1} \slashed{\nabla} \chi(X) \equiv
(\slashed{\del} + \frac{1}{4 \sqrt{2}} \Omega_{ABC} \Gamma^A \
\Gamma^B \ \Gamma^C) \chi(X)  - \frac{1}{\sqrt{2}} \Gamma^A
(U^{-1})^M_{\ A} \del_M \rho(Y) \chi(X). \ee

 Here, one might be puzzled that it is $\Omega_{ABC}$
rather than the $f_{ABC}$ and $\eta_A$ which appear in the reduced
Lagrangian. After all, it is $f_{ABC}$ and $\eta_A$ and not
$\Omega_{ABC}$, which are constrained to be constant by the
consistency requirement of the reduction of the NS-NS sector.
However, the following can be shown by using the commutation
relations in the Clifford algebra: \be \label{isom2} \frac{1}{4}
\Omega_{ABC} \Gamma^A \ \Gamma^B \ \Gamma^C \ \chi(X) =
\frac{1}{12} f_{ABC} \Gamma^A \ \Gamma^B \ \Gamma^C \ \chi(X)
-\frac{1}{2} f_B \Gamma^B \ \chi(X) \ee

\noindent Indeed \bea \frac{1}{4} \Omega_{ABC} \Gamma^A \Gamma^B
\Gamma^C & = & \frac{1}{12} \big(\Omega_{ABC} \Gamma^A \Gamma^B
\Gamma^C + \Omega_{BCA} \Gamma^B \Gamma^C \Gamma^A + \Omega_{CBA}
\Gamma^C
\Gamma^B \Gamma^A \big) \nonumber \\
& = & \frac{1}{12} \big(\Omega_{ABC} + \Omega_{BCA} +
\Omega_{CAB}) \Gamma^A \Gamma^B \Gamma^C  \nonumber \\
& & + \frac{1}{12} \big(\Omega_{BCA} (2 \eta^{AC} \Gamma^B - 2
\eta^{AB} \Gamma^C) + \Omega_{CAB} (2 \eta^{AC} \Gamma^B - 2
\eta^{CB} \Gamma^A) \big) \nonumber \\
& = & \frac{1}{12} f_{ABC} \Gamma^A \ \Gamma^B \ \Gamma^C
 -\frac{1}{2} f_B  \Gamma^B \eea where we have used the definitions (\ref{structure}) and (\ref{fa}) and the Clifford algebra identity
 \be \label{cliffid1} \Gamma^A \Gamma^B \Gamma^C  = \Gamma^B \Gamma^C \Gamma^A - 2
\eta^{AC} \Gamma^B + 2 \eta^{BA} \Gamma^C. \ee \noindent Plugging
this back in (\ref{hatchi1}), we get \bea \label{hatchi}
\slashed{\nabla} \chi(X) & \equiv & (\slashed{\del} + \frac{1}{12
\sqrt{2}} f_{ABC} \Gamma^A \ \Gamma^B \ \Gamma^C  + \frac{1}{2
\sqrt{2}} \eta_B \Gamma^B) \ \chi(X) \nonumber \\
& = & (\slashed{\del} + \frac{1}{6} f_{ABC} \psi^A \ \psi^B \
\psi^C + \frac{1}{2} \eta_B \psi^B) \ \chi(X). \eea

The Dirac operator  $\slashed{\nabla}$ is the same as the Dirac
operator introduced in  \cite{Geissbuhler2,Andriot11}, where they
study backgrounds with non-geometric  fluxes within the context of
flux formulation of DFT and $\beta$-supergravity, respectively,
without performing any duality twisted reduction (see also  the
associated papers \cite{Geissbuhler,Andriot12} and
\cite{parktwist}). It was shown in \cite{Andriot11} that the
Bianchi identities for the NS-NS fluxes are satisfied, only when
this Dirac operator is nilpotent. We will discuss this condition
of nilpotency further at the end of this subsection, when it
reappears as a condition to be satisfied for the gauge invariance
of the GDFT of the RR sector.

\subsubsection{Reduction of The Lagrangian}

The reduced Lagrangian can be obtained easily. If we plug
(\ref{redchi}) and (\ref{anzats}) in (\ref{mukaiactionfinal}) we
have \bea \label{actdef} e^{2 \rho(Y)} L_{{\rm red}} &=&
\frac{1}{4} \langle S \slashed{\nabla}\chi(X) , \ C^{-1}
(S^{-1})^\dagger \ciftS S^{-1}
S \slashed{\nabla}\chi(X) \rangle \nonumber \\
\label{deformed}  &=& \frac{1}{4} \langle
 \slashed{\nabla}\chi(X), \ C^{-1} \ciftS \slashed{\nabla}\chi(X) \rangle   \label{actiondeformed} \\
& = & \frac{1}{4} \langle
 \slashed{\partial}\chi(X)
, \ C^{-1} \ciftS \slashed{\partial}\chi(X)  \rangle \label{actiondeformed1} \\
&+& \frac{1}{4} \langle
  \bar{\chi}, \ C^{-1} \ciftS \slashed{\partial}\chi(X)  \rangle +  \frac{1}{4} \langle
 \slashed{\partial}\chi(X)
, \ C^{-1} \ciftS  \bar{\chi} \rangle \label{actiondeformed2} \\
&+& \frac{1}{4} \langle
 \bar{\chi}, \ C^{-1} \ciftS  \bar{\chi} \rangle, \label{actiondeformed3}
 \eea where $ \displaystyle \bar{\chi} = \frac{1}{12 \sqrt{2}}f_{ABC} \Gamma^A \Gamma^B
 \Gamma^C \chi + \frac{1}{2 \sqrt{2}} \eta_A \Gamma^A \chi$. Note that, we
 have used (\ref{daggerspin}) and (\ref{mukaiinvariant}) in
 passing from the first line to the second line.

 The term (\ref{actiondeformed1}) is the undeformed part of
 the Lagrangian.   The two terms in (\ref{actiondeformed2}) are
 equivalent as can be seen as follows\footnote{Note that
 $\bar{\chi}$ and $\chi$ have different chiralities. On the other
 hand, $C^{-1} \ciftS \bar{\chi}$ and $\bar{\chi}$ have the same chirality in 10 dimensions.
 Hence, $\slashed{\partial}\chi$ and $C^{-1} \ciftS \bar{\chi}$
 have the same chirality.}: \be \langle
  \bar{\chi}, \ C^{-1} \ciftS \slashed{\partial}\chi(X)  \rangle = -\langle
 C^{-1} \ciftS \bar{\chi}, C^{-1} \ciftS \ C^{-1} \ciftS \slashed{\partial}\chi(X)
 \rangle = -\langle
 C^{-1} \ciftS \bar{\chi},  \slashed{\partial}\chi(X)
 \rangle =  \langle
 \slashed{\partial}\chi(X)
, \ C^{-1} \ciftS  \bar{\chi} \rangle. \nonumber \ee Here we have
used the fact that $\cK = C^{-1} \ciftS \in Spin^-(10,10)$, which
explains the minus sign in applying (\ref{mukaiinvariant}) and
that $\cK^2 = 1$ and the Mukai pairing is skew-symmetric in 10
dimensions.  Now the two terms in (\ref{actiondeformed2}) add up
to give: \be \frac{1}{2} \langle
 \bar{\chi}, \ C^{-1} \ciftS  \slashed{\del}\chi \rangle = \frac{1}{2} \langle S_b
\bar{\chi}, \ S_b \ C^{-1} S_b^\dagger S_g^{-1} S_b
\slashed{\del}\chi \rangle = \frac{1}{2} \langle S_b \bar{\chi}, \
 C^{-1}  S_g^{-1} S_b \slashed{\del}\chi \rangle,
\ee where we have used (\ref{daggerspin}) and the invariance
property (\ref{mukaiinvariant}) along with the fact that $S_b \in
Spin^+(10,10)$. We have also plugged in the definition $ \ciftS=
S_b^\dagger S_g^{-1} S_b $. Note that
 here $b = b(X)$, $g = g(X)$ and $\chi = \chi(X)$, as all the $Y$ dependence in $\ciftS$ factorized
 out already in the first step. Also, requirement of constancy of $f_{ABC}$ and $\eta_A$ implies that $\bar{\chi} = \bar{\chi}(X)$.

 One can similarly compute the term
 (\ref{actiondeformed3}) and find that the reduced Lagrangian
 (\ref{deformed}) has the form
\be \label{reduced1} e^{2 \rho(Y)} L_{{\rm red}} = \frac{1}{4}
\langle  F(X),
 C^{-1}
S_g^{-1} F(X) \rangle + \frac{1}{2} \langle F(X),  C^{-1} S_g^{-1}
\bar{\chi}_B \rangle
 +  \frac{1}{4} \langle \bar{\chi}_B,   C^{-1}
S_g^{-1}
 \bar{\chi}_B \rangle. \ee

 Here we have defined $F(X)
= S_b \slashed{\del}\chi(X) =  e^{-B} \wedge
\slashed{\del}\chi(X)$ and $\bar{\chi}_B = S_b \bar{\chi} = e^{-B}
\wedge \bar{\chi}$. If the internal coordinates $X$ include no
dual coordinates, then this can be written as follows \be e^{2
\rho(Y)} L_{{\rm red}} = \frac{1}{4}   F(X) \wedge * F(X)
  + \frac{1}{2}  F(X) \wedge *
\bar{\chi}_B
 +  \frac{1}{4} \bar{\chi}_B \wedge *
 \bar{\chi}_B, \ee where $*$ is the Hodge duality operator with
 respect to the reduced metric $g(X)$.

On the other hand, the constraint reduces  to \be
\label{defduality} \slashed{\nabla} \chi(X) = - C^{-1} \ciftS \
\slashed{\nabla} \chi(X) \ee as can be shown easily by recalling
the definition $S^{\dagger} = C \tau(S) C^{-1}$ and the fact that
$\tau(S) = S^* = S^{-1}$ for $S \in Spin^+(n,n)$.

\subsubsection{Reduction of The Gauge Algebra}

 In order to find the gauge transformation rules for the
reduced theory  we plug the anzatse

\be \label{anzatsgauge} \xi^M(X, Y) = (U^{-1})^M_{\ A}
\hat{\xi}^A(X) \ee  \be \chi(X,Y) = e^{-\rho(Y)} S(Y) \chi(X) \ee
in the gauge transformation rules (\ref{gaugechi}),
(\ref{gaugelambda}) of the parent theory. This gives us the
following deformed gauge transformations for the spinor field
$\chi$:

\bea \label{defgaugexsi}  \delta_\xi \chi \ &=& \xi^M
\partial_M   (e^{-\rho(Y)} S(Y)  \chi ) \ + \ {1\over 2} \,  \partial_N \xi_M \Gamma^N
\Gamma^M (e^{-\rho(Y)} S(Y)  \chi ) \\
& =&(U^{-1})^M_{\ A} \hat{\xi}^A \del_M ( e^{-\rho(Y)} S \chi ) +
\frac{1}{2} \del_M (U^A_{\ N} \hat{\xi}_A ) \Gamma^M \Gamma^N (
e^{-\rho(Y)} S
\chi)  \nonumber \\
& = & e^{- \rho(Y)} S(Y) \big\{ \hat{\xi}^A \del_A  - \hat{\xi}^A
(U^{-1})^M_{\ A} \del_M \rho(Y) + \hat{\xi}^A (U^{-1})^M_{\ A}
(S^{-1} \del_M S) \nonumber \\ & & +  \frac{1}{2} U^A_{\ N} \del_M
\hat{\xi}_A (U^{-1})^M_{\ B} (U^{-1})^N_{\ C} \Gamma^B \Gamma^C  +
\frac{1}{2} (\del_M U^A_{\ N}) \hat{\xi}^A (U^{-1})^M_{\ B}
(U^{-1})^N_{\ C} \Gamma^B \Gamma^C
\big\} \chi \nonumber \\
& = & e^{- \rho(Y)} S(Y) \big\{ \hat{\xi}^A \del_A  + \frac{1}{2}
\del_B \hat{\xi}_C \Gamma^B \Gamma^C - \hat{\xi}^A (U^{-1})^M_{\
A} \del_M \rho(Y)\nonumber \\ & & + (\frac{1}{4}
\Omega^A_{\ \ BC} -\frac{1}{2} \Omega_{B \ \ C}^{\ \ A}) \Gamma^B \Gamma^C \big \} \chi \nonumber \\
& = & e^{- \rho(Y)} S(Y) \big\{ \hat{\xi}^A \del_A  + \frac{1}{2}
\del_B \hat{\xi}_C \Gamma^B \Gamma^C  + \frac{1}{4}( \Omega^A_{\ \
BC} - \Omega_{B \ \ C}^{\ \ A} - \Omega_{C \ \ B}^{\ \
 A} ) \hat{\xi}_A \Gamma^B \Gamma^C \nonumber \\
 & & -
\hat{\xi}^A (U^{-1})^M_{\ A} \del_M \rho(Y) - \frac{1}{2}
\Omega_{C \ \
B}^{\ \ A} \eta^{BC} \hat{\xi}_A \big\} \chi  \nonumber \\
& = &  e^{- \rho(Y)} S(Y) \big\{\delta_{\hat{\xi}} \chi +
\frac{1}{4} f^A_{\ \ BC} \ \hat{\xi}_A \ \Gamma^B \Gamma^C \chi +
\frac{1}{2} \eta^A \
\hat{\xi}_A \chi \big\} \nonumber \\
 & = & e^{- \rho(Y)} S(Y) \big\{ \hat{\delta}_{\hat{\xi}} \chi
\big \}
  \eea where we have used (\ref{cond1}), (\ref{structure}), (\ref{omega}), (\ref{esas}), (\ref{isom}) and Clifford algebra identities.
  In the last two lines we made the following
definitions: \bea \label{defgauge} \hat{\delta}_{\hat{\xi}} \chi &
\equiv &\delta_{\hat{\xi}} \chi + \frac{1}{4} f^A_{\ \ BC} \
\hat{\xi}_A \ \Gamma^B \Gamma^C \chi + \frac{1}{2} \eta^A \
\hat{\xi}_A \chi  \\
\delta_{\hat{\xi}} \chi & \equiv & \hat{\xi}^A \del_A \chi +
\frac{1}{2} \del_B \hat{\xi}_C \Gamma^B \Gamma^C \chi
 \eea


\noindent On the other hand, the deformation of the gauge
transformation (\ref{gaugelambda}) is found by plugging in the
anzats \be \label{anzatsgaugel} \lambda(X, Y) = e^{-\rho(Y)} S(Y)
\hl(X), \ee which then gives \bea \label{defgaugelambda}
\delta_{\lambda} \chi & = &
\slashed\del \big( e^{-\rho(Y)} S(Y) \hat{\lambda}\big) \\
& = & \frac{e^{-\rho(Y)}}{\sqrt{2}} \big( -\Gamma^M \del_M \rho(Y)
S(Y) \hat{\lambda} + \Gamma^M (\del_M S) \hl +
\Gamma^M S \del_M \hl \big) \nonumber \\
& = &  \frac{e^{-\rho(Y)}}{\sqrt{2}} \Gamma^M S \big(-  \del_M \rho(Y) + \del_M + S^{-1} \del_M S \big) \hl \nonumber \\
& = &  \frac{e^{-\rho(Y)}}{\sqrt{2}} S (U^{-1})^M_{\ A} \Gamma^A
\big(-  \del_M \rho(Y)+ \del_M + S^{-1} \del_M S
\big) \hl \nonumber \\
& = & \frac{e^{-\rho(Y)}}{\sqrt{2}} S \big( \Gamma^A \del_A +
\frac{1}{4} \Omega_{ABC} \Gamma^A
\Gamma^B \Gamma^C - (U^{-1})^M_{\ A} \del_M \rho(Y) \big) \hl \nonumber \\
& = &  e^{-\rho(Y)} S(Y) \big(\slashed\del + \frac{1}{12\sqrt{2}}
f_{ABC} \Gamma^A
\Gamma^B \Gamma^C + \frac{1}{2\sqrt{2}} \eta_{A} \Gamma^A \big)  \hl \nonumber \\
& = & e^{-\rho(Y)} S(Y) \big(\slashed\nabla \hl \big) \equiv
e^{-\rho(Y)} S(Y) \big(\hat{\delta}_{\hat\lambda} \chi \big). \eea

\subsubsection{Consistency of the Reduced Theory} Now that we have the
deformed action and the deformed gauge transformation rules, we
can analyze the conditions under which the GDFT of the RR sector
is consistent. Consistency is achieved if
\begin{itemize}
    \item[1.] $Y$ dependence drops  both in the reduced action and the gauge algebra.
    \item[2.] The reduced action is invariant under the deformed gauge
    transformation rules.
    \item[3.] The gauge algebra closes.
\end{itemize}
One can show that the constraints that arise from the consistency
of the reduction of the DFT of the NS-NS sector, that is, the
constancy of the fluxes $f_{ABC}$ and $\eta_A$ and the conditions
(\ref{cond1}-\ref{condlambda}) and
(\ref{onemlikosul}-\ref{orthogonality}) are sufficient to satisfy
the first and third items in the list above. When these conditions
are satisfied, the deformed gauge transformations we found above
close to form a gauge algebra as follows: \bea [\hdelta_{\hx_1},
\hdelta_{\hx_2} ] \chi & = & -
\hdelta_{[\hx_1, \hx_2 ]_f} \ \chi \\
& = & -[\hx_1, \hx_2 ]_f^A \del_A \chi - \frac{1}{2} \del_B
[\hx_1, \hx_2 ]_{f C} \Gamma^B \Gamma^C \chi - \frac{1}{4} f^A_{\
BC} [\hx_1, \hx_2 ]_{f A} \Gamma^B \Gamma^C \chi + \frac{1}{2} f^A
[\hx_1, \hx_2 ]_{f A}  \chi \nonumber \eea where \be [\hx_1, \hx_2
]_f^A \equiv [\hx_1, \hx_2 ]_C^A - f^A_{\ BC} \hx_1^B \hx_2^C, \ee
and  \be \big[\hdelta_{\hl}, \hdelta_{\hx} \big] =
\hdelta_{\hat{\cL}_{\hx} \hl} \chi \ee where \be \hat{\cL}_{\hx}
\hl = \hx^A \del_A \hl + \frac{1}{2} \del_B \hx_C \Gamma^B
\Gamma^C \hl + \frac{1}{4} f^A_{\ BC} \hx_A \Gamma^B \Gamma^C \hl
+ \frac{1}{2} \eta^A \hx_A \hl, \ee as can be verified by a
tedious calculation. In addition to the conditions
(\ref{cond1}-\ref{condlambda}) and
(\ref{onemlikosul}-\ref{orthogonality}), one also needs the
Clifford algebra identity (\ref{cliffid1}) and the following two
identities: \bea (S^{-1} \del_A
S ) (S^{-1} \del_B S )& = &- (\del_A S^{-1}) (\del_B S) \\
\Gamma^A \Gamma^B \Gamma^C \Gamma^D & = & \Gamma^C \Gamma^D
\Gamma^A \Gamma^B + 2 \eta^{CB} \Gamma^A \Gamma^D -  2 \eta^{DB}
\Gamma^A \Gamma^C +  2 \eta^{AC} \Gamma^D \Gamma^B -  2 \eta^{DA}
\Gamma^C \Gamma^B \nonumber \eea The last  identity follows
directly from the Clifford algebra.

The requirement of  gauge invariance of the GDFT action of the RR
sector imposes one more constraint on the fluxes.   Recall that
the DFT analogue of the p-form gauge transformation of the RR
fields has been deformed as \be \hat{\delta}_{\hat\lambda} \chi =
\slashed\nabla \hl. \ee Then we have \be
\hat{\delta}_{\hat\lambda} \slashed\nabla \chi = \slashed\nabla^2
\hl. \ee Therefore, the reduced action (\ref{actdef}) is invariant
under these gauge transformations if and only if the Dirac
operator $\slashed\nabla$ is nilpotent, that is, $\slashed\nabla^2
= 0$. As we mentioned above, this condition of nilpotency has
already appeared in \cite{Andriot11}. Let us now work out the
square of the Dirac operator. One can show that \bea 2
\slashed{\nabla}^2 \chi &=& \del_A \del^A \chi + \frac{1}{2}
f_{ABC} \Gamma^A \Gamma^B \del^C \chi - \eta_A \del^A \chi
-\frac{1}{4} f_{ABC} \eta^C \Gamma^A \Gamma^B \chi \nonumber \\
& & +\frac{1}{4} \eta_A \eta^A \chi + \frac{1}{16} f_{ABC}
f_{DE}^{\ \ \ A} \Gamma^D \Gamma^E \Gamma^B \Gamma^C \chi.  \eea
When the conditions
 (\ref{onemlikosul},\ref{onemlikosul2},\ref{external},\ref{orthogonality}) are
satisfied, the first line of the above expression vanishes. On the
other hand, applying the Jacobi identity (\ref{jacobi}), one can
show that the last term of the second line can be rewritten as
\bea \frac{1}{16} f_{ABC} f_{DE}^{\ \ \ A} \Gamma^D \Gamma^E
\Gamma^B \Gamma^C \chi &=& -\frac{1}{8} f_{ABC} f^{ABC} \chi
-\frac{1}{8} f_{ABC} f^{AD}_{\ \ \ D} \Gamma^B \Gamma^C \chi
\nonumber \\
& = & -\frac{1}{8} f_{ABC} f^{ABC} \chi, \eea where one uses in
passing to the second line the fact that $ f^{AD}_{\ \ \ D} = 0$.
 Then, we conclude that, up to the
constraints that are required for the consistency of the GDFT of
the NS-NS sector, we have \be \label{extraconstraint1}
\slashed{\nabla}^2 \chi = \big(2 \eta_A \eta^A - f_{ABC} f^{ABC}
\big) \chi = 0. \ee

Now let us consider the  gauge invariance of the deformed action
(\ref{deformed}) under the deformed gauge transformations with
parameter $\hat{\xi}$, for which we need $\hat{\delta}_{\hat{\xi}}
\cK$ and $\hat{\delta}_{\hat{\xi}} \slashed\nabla \chi$. In order
to calculate the first, one first has to note that the anzats
(\ref{anzats}) implies the following anzats for $\cK = C^{-1}
\ciftS$: \be\label{anzatsK} \cK(X,Y) = S(Y) \cK(X) S^{-1}(Y). \ee
Then using similar steps to above in the calculation of the
deformed gauge transformations for $\chi$, one finds\footnote{One
also needs the following identity in the proof, which had not been
needed before: \be S^{-1} \Gamma^{MN} S = \Gamma^{AB}
(U^{-1})^M_{\ A} (U^{-1})^N_{\ B},\ee where $\Gamma^{MN}$ is
defined in (\ref{spingen}).} \be \hat{\delta}_{\hat{\xi}}  \cK =
\hat{\xi}^A \del_A \cK + \frac{1}{2} [\Gamma^{AB}, \cK]
\big(\del_A \hat{\xi}_B + \frac{1}{2} f_{CAB} \hat{\xi}^C
\big).\ee On the other hand, one can compute \be
\hat{\delta}_{\hat{\xi}} (\slashed\nabla \chi) = \hat{\xi}^A
\del_A (\slashed\nabla \chi) + \big\{\frac{1}{2} \del_B
\hat{\xi}_C \Gamma^B \Gamma^C + \frac{1}{4} f^B_{\ CD} \hat{\xi}_B
\Gamma^C \Gamma^D + \frac{1}{2} \eta^B \hat{\xi}_B \big\}
\slashed\nabla \chi. \ee Plugging these in (\ref{deformed}) one
finds,\footnote{The details are similar to those in section 4.2.2
of \cite{dftRR}.} only using the constraints
(\ref{cond1}-\ref{condlambda}) and
(\ref{onemlikosul}-\ref{orthogonality}) \be
\hat{\delta}_{\hat{\xi}} L_{{\rm def}} = \hat{\xi}^A \del_A
L_{{\rm def}} + \del_A \hat{\xi}^A L_{{\rm def}} + \eta^A
\hat{\xi}_A L_{{\rm def}}. \ee Therefore, the deformed Lagrangian
is gauge invariant only when the fluxes $\eta^A $
vanish.\footnote{It has already been noted in \cite{Grana} that
these fluxes should vanish for the gauge invariance of the GDFT of
the NS-NS sector. It was also pointed out that this can be
circumvented by considering a modified reduction anzats, that
involveas a warp factor, as in \cite{Geissbuhler}. It would be
interesting to see whether the GDFT of the RR sector would remain
gauge invariant also for non-vanishing $\eta$ fluxes, by
introducing such a warp factor.} Combined with
(\ref{extraconstraint1}), following from the requirement of
nilpotency of the Dirac operator, we conclude that the requirement
of gauge invariance of the GDFT of the RR sector brings in the
extra condition \be \label{extraconstraint} f_{ABC} f^{ABC} = 0.
\ee The necessity of this extra constraint in the presence of RR
fields had already been anticipated in
\cite{Geissbuhler,Geissbuhler2} and had been verified by the
analysis of \cite{parktwist}. As was mentioned in section
(\ref{introduction}), the constraints of the GDFT (of the NS-NS
sector) are in one-to-one correspondence with the constraints of
half-maximal gauged supergravities. This extra condition we have
found implies that the gauged theory in hand corresponds to a
truncation of  maximal supergavity \cite{Aldazabal:2011yz}. We
also note that the gauge invariance of the duality relations
(\ref{defduality}) can also be verified easily, and does not
impose any extra constraints.

Before we finish, let us also comment on a possible modification
of the anzats (\ref{anzatschi2}), which introduces gaugings
associated with non-trivial RR fluxes. Note that the DFT action of
the RR sector (\ref{DFTaction}) is invariant under the global
shift symmetry $\chi \rightarrow \chi + \alpha $, which would make
it possible to introduce an anzats of the form $\chi(X, Y) =
\chi(X) + \tilde{\alpha}(Y)$. However, the gauge transformation
rules (\ref{gaugechi}) has an explicit dependence on $\chi$, which
then means that the  $Y$ dependence arising from such an anzats
would not drop  from  the reduced gauge transformation
rules\footnote{We also note that  it is possible to modify the
gauge transformation rules so as to be invariant under the global
shift as is done in \cite{HohmKwak}.}. One can still consider
introducing such an anzats by choosing the spinor field
$\tilde{\alpha}(Y)$ appropriately. Indeed, one can take \be
\tilde{\alpha}(Y) = e^{-\rho(Y)} S(Y) \alpha, \ee where $\alpha$
is a constant spinor field.  Then, the anzats (\ref{anzatschi2})
can be modified to \be \label{anzatschi3} \chi(X,Y) = e^{-\rho(Y)}
S(Y) (\chi(X) + \alpha). \ee When combined with the anzats
(\ref{anzats}), (\ref{anzatsdilaton}), (\ref{anzatsgauge}) and
(\ref{anzatsgaugel}), a reduction with the modified anzats
(\ref{anzatschi3}) leads to a consistent theory, for which the $Y$
dependence drops  from the action and the gauge transformations.
Plugging (\ref{anzatschi3}) in $\Gamma^M \del_M \chi(X,Y)$, one
finds (we now take $\eta^A = 0$) \bea \slashed\del \chi(X,Y) & = &
\frac{1}{\sqrt{2}} \Gamma^M \del_M \big(e^{-\rho(Y)} S(Y) \chi(X)
\big) + \frac{1}{\sqrt{2}} \Gamma^M
\del_M \big(e^{-\rho(Y)} S(Y) \alpha \big) \nonumber \\
& = & e^{-\rho(Y)} S(Y) \slashed\nabla \chi(X) +
\frac{1}{\sqrt{2}} e^{-\rho(Y)} S(Y)
(U^{-1})^M_{\ A} \Gamma^A \big(S^{-1} \del_M S) \alpha \nonumber \\
& = & e^{-\rho(Y)} S(Y) \big\{\slashed\nabla \chi(X) +
\frac{1}{12\sqrt{2}} f_{ABC} \Gamma^A \Gamma^B \Gamma^C
\alpha\big\}, \eea where we have used the identities (\ref{esas}),
(\ref{isom}) and (\ref{isom2}) and $\slashed\nabla$ is as
before.\footnote{Note that we have not taken the terms associated
with the derivative of $\rho$ into account, as they combine with
$f_A$ in (\ref{isom2}) to give $\eta_A$, which we take zero now.}
The reduced theory in this case is \be \label{modlag} L_{{\rm
red}} = \frac{1}{4} \big\langle \slashed\nabla \chi +
\frac{1}{12\sqrt{2}} f_{ABC} \Gamma^A \Gamma^B \Gamma^C \alpha, \
\cK \big( \slashed\nabla \chi + \frac{1}{12\sqrt{2}} f_{ABC}
\Gamma^A \Gamma^B \Gamma^C \alpha \big) \big\rangle. \ee The
Lagrangian (\ref{modlag})  has to be supplemented by the following
duality relation \be \slashed\nabla \chi + \frac{1}{12\sqrt{2}}
f_{ABC} \Gamma^A \Gamma^B \Gamma^C \alpha = - \cK \big(
\slashed\nabla \chi + \frac{1}{12\sqrt{2}} f_{ABC} \Gamma^A
\Gamma^B \Gamma^C \alpha \big). \ee The Lagrangian and the duality
relation is invariant under the following gauge transformation \be
\hat{\delta}_{\hat{\xi}} \chi = \hat{\xi}^A \del_A \chi +
\frac{1}{2} \del_B \hat{\xi}_C \Gamma^B \Gamma^C \big( \chi +
\alpha \big) + \frac{1}{4} f^A_{\ \ BC} \ \hat{\xi}_A \ \Gamma^B
\Gamma^C \big( \chi + \alpha \big).
 \ee


\section{Conclusions and Outlook}\label{conc}

In this paper, we studied the duality twisted reduction of the
Double Field Theory  of the RR sector of massless Type II theory.
This sector of DFT has a global $Spin^+(n,n)$ symmetry, which we
have utilized to introduce the duality twisted anzats. We obtained
the reduced action and the gauge transformation rules and showed
that the gauge algebra closes. The fact that the Lie algebras of
$Spin(n,n)$ and $SO(n,n)$ are isomorphic plays a crucial role in
our analysis.

Our reduction anzats is determined by a $Spin^+(n,n)$ element $S$.
Under the double covering homomorphism $\rho$ between $Spin(n,n)$
and $SO(n,n)$, the twist element $S$ projects to an element  $U
\in SO^+(n,n)$. This then implies a duality twist in the
accompanying reduction of the NS-NS sector of DFT, through the
matrix $U$. The duality twisted reduction of the NS-NS sector has
already been studied by several groups
\cite{Grana,Geissbuhler,Aldazabal}. As was shown in these works,
the consistency of the reduced theory and its gauge algebra places
restrictions on the fluxes determined by the twist matrix $U$. It
was shown in \cite{Grana} that these constraints are in one-to-one
correspondence with the constraints of half-maximal gauged
supergravity. All these constraints are also crucial for the
consistency of the GDFT of the RR sector. In addition, we have
shown here that the requirement of gauge invariance in the RR
sector imposes the extra constraint (\ref{extraconstraint}), which
also appeared in \cite{parktwist}. It is known that any
half-maximal gauged supergravity that satisfies this constraint
can be uplifted to maximal gauged supergavity
\cite{Aldazabal:2011yz}. Therefore, the existence of this extra
constraint   can be seen as a sign  that the reduction we have
studied here should be related to duality twisted reductions of
Exceptional Field Theory (EFT), which is a U-duality invariant
extension of supergravity \cite{HullMth, BermanMth, HohmExc}.
Indeed, the reduction of EFT on generalised parallelisable
manifolds \cite{leewaldram} (which corresponds to a reduction with
a duality twisted anzats of the type we have considered here)
gives rise to maximal gauged supergravity upon imposing a section
constraint, which is the analogue of the strong constraint of DFT
\cite{bermanSS,aldazabalSS,samtlebenSS}. A flux formulation of (a
particular type of) EFT is also available and geometric and
non-geometric RR fluxes were studied also in this formulation
\cite{malekblair}. For recent work on how to truncate such
theories further to half-maximal gauged supergravities, see
\cite{malek1, malek2}.

An interesting feature of our reduced action is the natural
appearance of the nilpotent Dirac operator (\ref{hatchi}),
associated with the spinorial covariant derivative acting in the
RR sector. This Dirac operator has already appeared in various
papers before (e.g.\cite{Geissbuhler2,Andriot11,parktwist}). It
was shown in \cite{Andriot11} that the Bianchi identities for the
NS-NS fluxes are satisfied, only when this Dirac operator is
nilpotent and the same condition arises here from the analysis of
the gauge invariance of the GDFT of the RR sector.  Note that the
flux dependent terms in the Dirac operator involves (products of)
Gamma matrices. As we discussed in section (\ref{math}), the
spinorial action of these Clifford algebra elements on the spinor
field $\chi$ (which is equivalently a differential form) is by
contraction, when they belong to the orthogonal complement of the
vector subspace, whose exterior algebra carries the spinorial
representation of the Clifford algebra. In other words, the Gamma
matrices with a lower index act on the spinor fields  by
contraction. For certain choices of twists, this gives the
possibility of inducing 0-forms as deformation terms in the
reduced action.  We will explore this feature in \cite{Aybike},
where we study massive deformations of Type IIA theory within DFT.

In analyzing the reduction of the DFT Lagrangian of the RR sector,
we found it useful to rewrite it in terms of Mukai pairing, which
is a $Spin(n,n)$ invariant bilinear form on the space of spinors.
We believe that the Lagrangian, when written in the form
(\ref{mukaiactionfinal})  is worth further study. Note that
(\ref{mukaiactionfinal}) gives a non-vanishing $n$-form, which is
a volume form when the underlying manifold $M$ is $n$ dimensional
as in generalized geometry of Hitchin. As a Lagrangian for DFT, it
gives us an $n$-form on the $2n$ dimensional doubled manifold.
However, it is a very special $n$-form. Recall that the spinor
field $\chi$ is a section of the restricted exterior bundle
$\bigwedge^\bullet T^*M^{{\rm doub}}$, in the sense that at each
point the cotangent space is restricted to a maximally isotropic
subspace with respect to the metric $\eta$. This then implies that
the $n$-form produced by the Lagrangian (\ref{mukaiactionfinal})
belongs to a 1-dimensional subspace of the $(2n)! / (n)! (n)! $
dimensional space of all possible $n$-forms on a $2n$ dimensional
manifold, as it can have components only along these $n$
restricted directions. (Note that the value of the form still
depends on the $\tilde{x}$ coordinates of the manifold). Then, one
can naturally identify this $n$-form with a scalar function (a
0-form),  which then becomes the Lagrangian density to be
integrated on the whole doubled manifold. It would be desirable to
come up with a Lagrangian which produces a volume form for the
whole doubled manifold. This obviously calls for a better
understanding of the differential geometric features of doubled
manifolds.  We believe that this direction deserves further study
and we hope to come back to these issues  in future work.

\section*{Acknowledgments}
This work is supported by the Turkish Council of Research and
Technology (T\"{U}B\.{I}TAK) through the ARDEB 1001  project with
grant number 114F321, in conjunction with the COST action MP1405
QSPACE. I would like to thank Dublin Institute for Advanced
Studies, where writing of this work was finalized, for
hospitality, and Chris Hull for useful discussions.

\end{document}